\documentclass[universe,review,accept,pdftex,oneauthor]{Definitions/mdpi} 
\firstpage{1} 
\pubvolume{1}
\issuenum{1}
\articlenumber{0}
\pubyear{2026}
\copyrightyear{2026}
\externaleditor{Firstname Lastname} 
\datereceived{30 June 2026 } 
\daterevised{9 August 2026 } 
\dateaccepted{13 August 2026 } 
\datepublished{ } 
\hreflink{https://doi.org/} 

\usepackage{pgfplots}
\usepackage{natbib}
\usepackage{tabularx}

\newcommand{\apj}{ Astrophys. J.}
\newcommand{\apjs}{ Astrophys. J. Suppl. Ser.}
\newcommand{\apjl}{Astrophys. J.}
\newcommand{\aap}{Astron. Astrophys.}
\newcommand{\aapr}{ Astron. Astrophys. Rev.}

\newcommand{\mnras}{Mon. Not. R. Astron. Soc.}
\newcommand{\araa}{Annu. Rev. Astron. Astrophys.}
\newcommand{\jcap}{J. Cosmol. Astropart. Phys.}
\newcommand{\nat}{Nature}
\newcommand{\physrep}{Phys. Rep.}
\newcommand{\prd}{Phys. Rev. D}
\newcommand{\prl}{Phys. Rev. Lett.}
\newcommand{\ssr}{Space Sci. Rev.}

\newcommand{\rmo}[1]{#1}

\Title{Merging Galaxy Clusters and the Search for New Physics of Dark Matter: A Review}

\Author{Rogério Monteiro-Oliveira$^{1,2}$\orcidA{}}

\AuthorNames{Rogério Monteiro-Oliveira}

\address{%
$^{1}$ \quad Observatório Nacional (ON/MCTI), Rua General José Cristino, 77, Rio de Janeiro 20921-400, Brazil; rmo@on.br \\
$^{2}$ \quad Institute of Astronomy and Astrophysics, Academia Sinica, Taipei 106319, Taiwan}

\abstract{Merging galaxy clusters represent one of the most powerful macroscopic laboratories in the Universe for searching for new physics within the dark sector. High-velocity cosmic collisions inherently separate the dark matter and stellar components from the highly collisional, X-ray-emitting intracluster gas. These massive systems provide an ideal environment to probe the fundamental nature of dark matter, specifically testing whether it behaves as a strictly collisionless particle or exhibits non-zero self-interactions. While pioneering systems like the Bullet Cluster historically demonstrated the macroscopic decoupling of dark and ordinary matter, the field has evolved into a sophisticated discipline driven by multi-disciplinary methodologies. This review synthesizes recent theoretical and empirical advances in interpreting post-collision dynamics. It examines how the synergy of combined approaches---integrating multi-wavelength observations from gravitational lensing and X-ray mapping with high-fidelity $N$-body hydrodynamical simulations---allows the translation of macroscopic spatial observables into stringent constraints on microscopic particle properties. Through this synthesis, the work evaluates how leveraging heterogeneous merger ensembles can reliably advance the ongoing search for physics beyond the standard cosmological model.}


\begin{document}


\section{Introduction}
\label{sec:intro}

Dark matter serves as the foundational architect of large-scale structure in the Universe. As~the bedrock of the standard cosmological paradigm ($\Lambda$CDM), it provides the essential gravitational scaffolding required for cosmic evolution. Because~it is immune to the immense radiation pressure that impeded baryonic collapse prior to the epoch of recombination, cold dark matter (CDM) was able to gravitationally decouple and establish the primordial potential wells into which ordinary matter later collapsed \citep{Blumenthal84, Planck20}. While alternative theoretical frameworks, ranging from Modified Newtonian Dynamics (MOND) and emergent gravity to Hot Dark Matter scenarios driven by massive neutrinos, attempt to circumvent the need for a novel particle, they persistently struggle to simultaneously reproduce the primordial power spectrum and the non-linear growth of cosmic structures \citep{White83,Clowe06}. In~contrast, a~modern generation of high-resolution N-body and fully coupled hydrodynamical simulations grounded in $\Lambda$CDM yields astonishing agreement with a diverse array of observational probes. From~the precise morphological acoustic peaks of the Cosmic Microwave Background (CMB) to the statistical clustering of galaxies and the intricate topology of the cosmic web, these multi-scale successes unequivocally cement dark matter’s indispensable role in driving the dynamical evolution of the Universe \citep{Springel05,Vogelsberger14,Alam17}.

The CDM exerts a widespread and measurable gravitational influence across a vast range of cosmic scales, providing a unifying theoretical framework for diverse astrophysical phenomena. On~galactic scales, some of the most compelling evidence comes from the extended, flat rotation curves of disk galaxies. These velocity measurements require the presence of a massive, invisible dark matter halo extending far beyond the luminous optical disk to prevent the expected Keplerian drop-off \citep{Rubin80}. Moving to the scale of galaxy clusters, multi-wavelength observations consistently demonstrate that normal baryonic matter accounts for only a small fraction of the total mass. This conclusion is independently supported by the velocity dispersions of member galaxies \citep{Zwicky33}, the~hydrostatic equilibrium of the hot, X-ray-emitting intracluster medium (ICM) \citep{Vikhlinin06}, and~the distortion of background light through gravitational lensing \citep{Hoekstra13}. On~the largest cosmological scales, the~temperature and polarization fluctuations of the CMB require a collisionless dark matter component to accurately reproduce the relative heights of the acoustic peaks \citep{Aiola20}. This highly consistent picture, spanning from the dynamics of individual galactic halos to the large-scale structure of the cosmic web, solidifies dark matter as an essential cornerstone of the modern cosmological model \citep{Frenk12}.

Despite this overwhelming macroscopic evidence, the~microscopic identity of dark matter remains one of the most profound unresolved puzzles in fundamental physics~\citep{Feng10}. The~standard $\Lambda$CDM framework operates on the phenomenological assumption that dark matter is a strictly cold and collisionless fluid, interacting with the Standard Model primarily (and perhaps exclusively) via gravity~\citep{Bertone05, Frenk12}. For~decades, theoretical efforts have been heavily anchored by Weakly Interacting Massive Particles (WIMPs), driven by the remarkable theoretical coincidence that particles with electroweak-scale cross-sections naturally yield the observed relic density through thermal freeze-out in the early Universe~\citep{Steigman85, Jungman96}. However, the~persistent absence of definitive signals from deep-underground direct detection experiments, terrestrial particle colliders, and~indirect cosmic-ray searches has prompted a critical re-evaluation of the dark sector's potential complexity \citep{Arcadi18, Boveia18, Schumann19}.

Consequently, the~astrophysical arena has become the paramount frontier for constraining dark matter's underlying particle nature \citep{Buckley18}. If~the dark sector encompasses richer dynamics---such as non-trivial self-interactions, ultra-light quantum wave behavior, or~complex formation histories---these fundamental particle properties will inevitably leave macroscopic imprints on the assembly and internal structure of cosmic halos \citep{Hui17, Tulin18}. By~shifting the focus from terrestrial laboratories to cosmic environments, the~nonlinear regime of gravitational collapse can be leveraged to probe interaction cross-sections, masses, and~decay rates that are fundamentally inaccessible to Earth-bound experiments \citep{Bauer15, Zavala19}.

\begin{figure}[H]
\includegraphics[width=\textwidth]{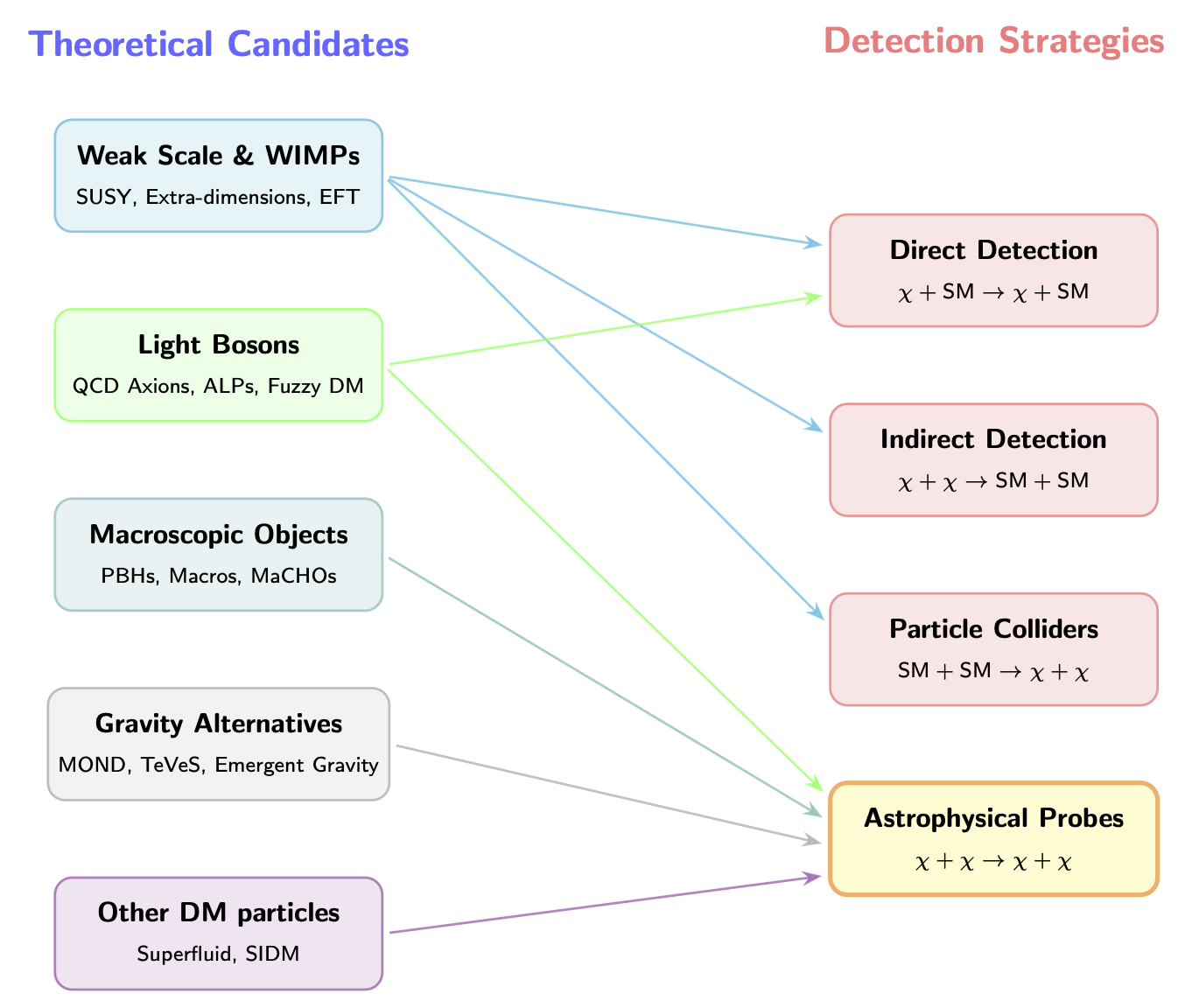}
\caption{Theoretical taxonomy and empirical verification strategies for the dark sector. The~theoretical landscape (\textbf{left}) categorizes dark matter candidates and alternative frameworks by physical mechanism, spanning from ultralight bosons and canonical weak-scale WIMPs to macroscopic objects and modified gravitational theories. These models are constrained by four complementary detection frontiers (\textbf{right}), which track potential non-gravitational couplings or dark-sector dynamics. The~reaction equations, where $\chi$ denotes the generic dark matter particle and SM represents Standard Model components, illustrate the fundamental scattering, annihilation, or~production channels for each approach. Connecting lines map the theoretical candidates to the experimental methodologies that most stringently probe them. While terrestrial colliders and direct/indirect detection primarily test fundamental SM interactions, macroscopic frameworks and ``other dark particles'' (such as SIDM) necessitate fundamentally different observational strategies. Consequently, astrophysical probes serve as the paramount, and~often exclusive, laboratories for these~models.} 
\label{fig:dm_synthesis}
\end{figure}

To explore these richer dynamics and bridge the gap between microscopic particle physics and macroscopic observable structures, several alternative dark matter frameworks have been proposed (see Figure~\ref{fig:dm_synthesis}):

\begin{itemize}

\item Self-Interacting Dark Matter (SIDM): Introduces a non-negligible scattering cross-section between dark matter particles. This allows the dark matter fluid to exchange energy and momentum, effectively thermalizing the dense central regions of halos and dynamically altering their phase-space evolution \citep{Spergel00}.

\item Warm Dark Matter (WDM): Imbues dark matter with a larger initial velocity dispersion, introducing a characteristic free-streaming length. This kinematically suppresses primordial density fluctuations below a specific mass threshold, altering the hierarchy of non-linear structure formation at sub-galactic scales \citep{Bode01}.

\item Fuzzy Dark Matter (FDM): Postulates ultralight bosonic particles ($\sim$$10^{-22}\text{ eV}$) where macroscopic quantum wave effects and inherent quantum pressure dominate the dynamics. This inherently smooths out small-scale clustering and generates localized, macroscopic soliton structures at the centers of dark matter halos \citep{Ferreira21}.

\item Standard Particle Candidates vs. Macroscopic Objects: While traditional cold, collisionless particle candidates like WIMPs and QCD axions remain deeply motivated by particle physics, accurately mapping their macroscopic distribution increasingly requires coupling them to complex, non-linear baryonic feedback mechanisms to account for observable phase-space modifications \citep{Bertone05, Pontzen12}. Conversely, macroscopic objects, such as primordial black holes (PBHs) and MaCHOs, eschew these subatomic particle interactions entirely, bypassing terrestrial detection frontiers and requiring observational strategies rooted purely in gravitational and astrophysical signatures~\citep{Carr21, Green21}.

\end{itemize}\vspace{-12pt}

\begin{figure}[H]
\includegraphics[width=0.99\textwidth]{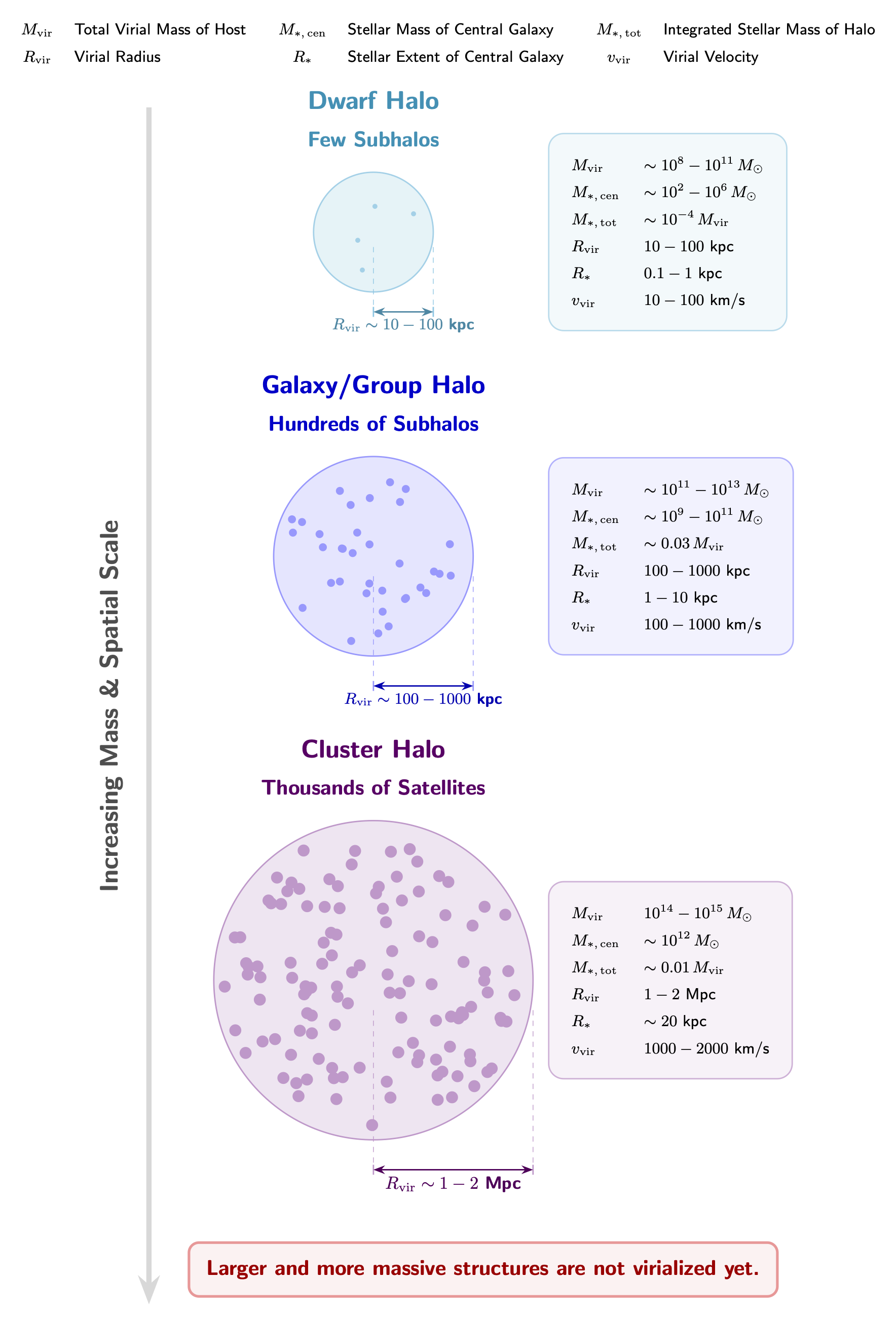}
\caption{Hierarchical scaling of dark matter halos as astrophysical laboratories. Evolving vertically from localized dwarf systems to massive galaxy clusters, these structures serve as natural laboratories whose macroscopic properties directly constrain underlying microphysical models. The~structural demographics, such as the substructure population (represented by internal particles) and spatial extent (projection markers, $R_{\rm vir}$), scale proportionally with their overarching host mass, increasing from kiloparsecs to megaparsecs. As~indicated by the threshold at the bottom of the hierarchy, galaxy clusters ($M_{\rm vir} \approx 10^{14}$--$10^{15} \, M_\odot$) represent the current limit of gravitational collapse; larger and more massive cosmic structures have not yet had sufficient time to reach~equilibrium.}
\label{fig:halo_scaling}
\end{figure}

Because each of these theoretical frameworks predicts distinct structural signatures, cosmic environments spanning a vast hierarchy of scales, from~dwarf spheroidal galaxies to massive galaxy clusters, serve as indispensable natural laboratories for mapping these macroscopic properties (see Figure~\ref{fig:halo_scaling}). As~emphasized across numerous modern reviews \citep{Einasto09,Bauer15,Buckley18,Tulin18,Boddy22,Chakrabarti22,Adhikari25}, isolating and characterizing the dynamics within these varied environments provides the critical empirical benchmarks needed to constrain the underlying dark matter particle models \citep{Wechsler18}. Specifically, the~measurable properties of these dark matter halos---such as their abundance, structural density profiles, and~phase-space distributions---directly reflect fundamental particle characteristics. This macro-micro connection proves particularly powerful for testing theoretical extensions beyond the standard WIMP paradigm \citep{Battaglieri17}. Crucially, astrophysical data uniquely constrain the parameter spaces of candidates that easily evade direct terrestrial detection, spanning from ultra-heavy primordial black holes \citep{Carr21} to ultralight axion-like particles \citep{Marsh16}. Ultimately, astrophysical probes do not merely complement direct laboratory detection efforts (Figure~\ref{fig:dm_synthesis}); they provide transformative and often exclusive insights into dark matter's fundamental physics by anchoring cosmological-scale dynamics to microphysical theory \citep{Bauer15}.

At the extreme of this mass spectrum, galaxy clusters yield the most stringent observational constraints on dark sector dynamics \citep{Allen11}. Representing the most massive gravitationally bound structures in the Universe ($M \approx 10^{14}$--$10^{15} \, M_\odot$), these systems furnish an unprecedented environment for interrogating the fundamental properties of dark matter, owing to their complex multiphase composition and extreme dynamical conditions~\citep{Kravtsov12}. Although~their total mass budget is heavily dominated by an extended dark matter halo ($\sim$$85\%$) that dictates the overarching kinematic potential, clusters are fundamentally tripartite systems \citep{Voit05}. They encapsulate a hot, diffuse ICM, which emits profusely in the X-ray regime and induces the Sunyaev--Zel'dovich (SZ) effect \citep{Carlstrom02}, alongside a discrete population of member galaxies whose phase-space distributions are optically trackable \citep{Girardi98a}. The~interplay among these distinct baryonic and dark components facilitates rigorous empirical tests of dark matter physics beyond the canonical collisionless paradigm. Specifically, highly energetic dynamic events, such as major cluster mergers, can induce quantifiable spatial offsets between the highly collisional ICM and the nominally collisionless dark matter and galactic components \citep{Markevitch04, Harvey15}. Furthermore, morphological deviations in central density profiles, coupled with the survivability statistics of orbiting subhalos, operate as highly sensitive probes for SIDM and alternative non-gravitational couplings \citep{Peter13, Tulin18}. Crucially, the~synthesis of weak and strong gravitational lensing with multi-wavelength datasets permits the reconstruction of the underlying dark matter potential independent of convoluted baryonic processes, such as active galactic nucleus (AGN) feedback and thermal shock heating \citep{Kneib11, Fabian12}. By~effectively disentangling these purely gravitational signatures from baryonic phenomena, galaxy clusters yield definitive constraints on the cross-sections of dark matter, effectively serving as macroscopic particle colliders capable of accessing interaction regimes thoroughly inaccessible to terrestrial experiments \citep{Massey15}.

Designed to bridge the gap between cosmological-scale observations and microphysical hypotheses, this review synthesizes recent theoretical and empirical breakthroughs in the study of merging galaxy clusters. Our primary objective is to evaluate these colossal systems not merely as astrophysical phenomena, but~as macroscopic particle colliders uniquely positioned to reveal the fundamental nature of dark matter. To~establish this framework, the~paper follows a logical trajectory connecting cosmic dynamics to particle interactions. We begin in Section~\ref{sec:structure_formation} by reviewing the foundational role of dark matter in driving structure formation across scales, before~exploring MOND as a major theoretical alternative in Section~\ref{sec:mond}. Section~\ref{sec:bullet_cluster} then revisits the Bullet Cluster, dissecting its legacy as the archetypal empirical proof of dark matter decoupling. Shifting focus to sub-galactic regimes, Section~\ref{sec:small_scale_crisis} addresses the persistent anomalies of the small-scale crisis that challenge purely collisionless models and motivate paradigms such as self-interacting dark matter (SIDM). This microphysical transition culminates in Section~\ref{sec:mergers_as_colliders}, which details how cluster mergers map dark sector scattering cross-sections. After~confronting the field's pressing observational and theoretical hurdles in Section~\ref{sec:challenges}, Section~\ref{sec:conclusions} outlines the conclusions of this~review.\vspace{-9pt}

\section{The Cosmological Framework: Dark Matter and Structure~Formation}
\label{sec:structure_formation}

The $\Lambda$CDM model has proven remarkably successful in explaining a wide range of observational phenomena across the history of the Universe~\cite{Abbott22,Dalal23,Wright25}. A~fundamental cornerstone of this empirical success is the inclusion of CDM, a~non-baryonic, collisionless, and~pressureless component that dominates the cosmic matter budget~\cite{Peebles82}. While the precise nature of the dark matter particle remains elusive, its macroscopic gravitational effects are consistently and unequivocally observed across all cosmological scales~\cite{Trimble87,Bertone18}. This section examines the theoretical frameworks and multi-scale observational evidence demonstrating that structure formation, from~primordial perturbations to the modern cosmic web, is impossible without the driving influence of dark matter~\cite{Mo10}.

\subsection{The Cosmic Microwave Background: Primordial Perturbations and Boundary~Conditions}
\label{sec:cmb_anisotropies}

The CMB provides a pristine snapshot of the Universe at the epoch of recombination ($z \sim 1100$), when photons dynamically decoupled from baryons and initiated free-streaming~\cite{Hu02}. The~temperature anisotropies mapped by the CMB encode a wealth of information regarding primordial perturbations, early plasma physics, and~the total energetic composition of the cosmic fluid~\cite{Durrer20}. Crucially, the~observed amplitude of these temperature fluctuations ($\Delta T / T \sim 10^{-5}$) is entirely insufficient to seed the formation of non-linear structures by the present day ($z=0$) in a baryon-only universe~\cite{Bond84}.

Baryons were prevented from undergoing gravitational collapse prior to recombination due to their tight coupling to the immense radiation pressure of the photon fluid~\cite{Silk68}. CDM, being pressureless and immune to electromagnetic interactions, was able to decouple much earlier. Once inside the horizon, CDM perturbations---quantified by the dimensionless density contrast $\delta = (\rho - \bar{\rho})/\bar{\rho}$---grew only logarithmically during the radiation-dominated era due to the severe Hubble drag~\cite{Meszaros74}. However, following matter-radiation equality ($z \sim 3400$), these perturbations were free to grow aggressively, scaling linearly with the cosmic scale factor $a$ ($\delta \propto a$) \cite{Dodelson20}. Consequently, dark matter established deep primordial gravitational potential wells, acting as a vital cosmological scaffolding, into~which ordinary baryons fell only after recombination~\cite{Hu96}.

This interplay is mathematically captured by the acoustic peaks in the angular power spectrum, which arise from acoustic oscillations in the photon-baryon fluid. The~physics of these oscillations is governed by the coupled Einstein--Boltzmann equations, where the evolution of the photon temperature monopole, $\Theta_0(k,\eta)$, in~the tight-coupling approximation is given by:
\begin{equation}
    \ddot{\Theta}_0 + \frac{\dot{R}}{1+R} \dot{\Theta}_0 + c_s^2 k^2 \Theta_0 = - \ddot{\Phi} - \frac{\dot{R}}{1+R} \dot{\Phi} - \frac{1}{3} k^2 \Psi,
\end{equation}
where $\eta$ is the conformal time, $R = \frac{3\rho_b}{4\rho_\gamma}$ dictates the baryon loading (i.e., the~ratio of baryon energy density $\rho_b$ to photon energy density $\rho_\gamma$), $c_s = \frac{1}{\sqrt{3(1+R)}}$ represents the fluid's sound speed, and~$\Phi$ and $\Psi$ denote the Newtonian gravitational potentials~\cite{Dodelson20}. Without~dark matter, these potentials would decay rapidly inside the horizon, suppressing the driving force of the acoustic oscillations and drastically altering the resulting peak structure~\cite{Seljak96}.

The morphology of these acoustic peaks yields exact cosmological boundaries. The~angular scale of the primary peak maps directly to a spatially flat Universe, whereas the contrast between the odd-numbered compression peaks and even-numbered rarefaction peaks dictates the baryon-to-dark-matter ratio; a dominant CDM component is essential to provide the gravitational wells that resist radiation pressure~\cite{Spergel03}. This early-Universe requirement is robustly quantified by Planck satellite data, which constrains the physical baryon density to $\Omega_{\rm b} h^2 = 0.0224 \pm 0.0001$ and the CDM density to $\Omega_{\rm CDM} h^2 = 0.120 \pm 0.001$, firmly cementing the inventory of the primordial plasma~\cite{Planck20}.

\subsection{The Linear Regime: The Matter Power Spectrum and Baryonic~Oscillations}
\label{sec:linear_growth}

The perturbations responsible for CMB anisotropies act as the direct progenitors of modern large-scale structure (LSS) \cite{Nelson19}. Post matter-radiation equality, the~evolution of the matter overdensity, $\delta_m$, in~the linear regime is dictated by:
\begin{equation}
    \ddot{\delta}_m + 2 H \dot{\delta}_m = 4\pi G \rho_m \delta_m,
\end{equation}
where $H$ is the Hubble parameter and $\rho_m$ encapsulates the total matter density~\cite{Peebles93}. In~the absence of dark matter, Silk damping would have catastrophically erased small-scale baryonic structures in the pre-recombination plasma~\cite{Silk68}. CDM preserves these small-scale modes, yielding a matter power spectrum that perfectly matches the observed statistical clustering of galaxies in massive, state-of-the-art redshift surveys, such as the final legacy sample of the Extended Baryon Oscillation Spectroscopic Survey (eBOSS,~\cite{Alam21}), and~the recent high-precision maps from the Dark Energy Spectroscopic Instrument (DESI,~\cite{Adame25}). The~characteristic turnover scale in this power spectrum corresponds precisely to the horizon size at matter-radiation equality, a~feature that cannot be replicated without invoking CDM~\cite{Planck20, Alam21}. Precision measurements of this horizon scale now provide some of the cleanest constraints on the total matter density of the Universe, independent of late-time acceleration physics~\cite{Ivanov20,Philcox22}.

Furthermore, Baryon Acoustic Oscillations (BAO) embed a standard physical ruler onto this spatial distribution. The~fundamental scale of these oscillations is determined by the comoving sound horizon at the drag epoch:
\begin{equation}
    r_d = \int_0^{t_{\rm drag}} \frac{c_s}{a(t)} dt,
\end{equation}
where $c_s$ is the aforementioned photon-baryon sound speed~\cite{Dodelson20, Planck20}. Following the drag epoch, baryons coalesce into the pre-existing CDM potentials, permanently freezing the BAO signature into the spatial correlation function of galaxies, a~feature confirmed with high statistical significance by modern LSS surveys~\cite{Alam21, Adame25}.

\subsection{Non-Linear Collapse and High-Redshift Structural~Assembly}
\label{sec:high_redshift}

The indispensable nature of dark matter is further emphasized by the observation of exceptionally massive galaxies and supermassive black holes ($M > 10^9 \, M_\odot$) at early cosmic epochs ($z > 6$) \cite{WangF21,Labbe23,Boylan-Kolchin23}. The~formation of such extreme systems necessitates the rapid assembly of deep gravitational potential wells almost immediately following recombination~\cite{Volonteri10}. The~theoretical framework for this non-linear collapse is described by the Press--Schechter formalism, which models the comoving number density of dark matter halos as:
\begin{equation}
    \frac{dn}{dM} \propto \left(\frac{\rho_m}{M^2}\right) \left|\frac{d\ln \sigma}{d\ln M}\right| \exp\left[-\frac{\delta_c^2}{2 \sigma^2(M)}\right],
\end{equation}
where $\delta_c$ represents the critical density threshold for spherical collapse and $\sigma(M)$ defines the variance of the smoothed density field~\cite{Press74}. CDM uniquely provides the necessary high-$\sigma$ structural peaks that collapse early enough to accommodate these high-redshift observations; without it, the~abbreviated cosmic time available severely precludes their formation~\cite{Springel18}.

\subsection{Galaxy Clusters: Lensing and Dynamical Mass~Probes}
\label{sec:clusters_lensing}

Occupying the extreme high-mass tail of this hierarchical assembly, galaxy clusters offer direct, multiphysics tests of the cosmic mass budget~\cite{Kravtsov12,Tam26}. Cluster masses are traditionally calculated via independent methodologies: the kinematic velocity dispersion of member galaxies and the Jeans equation~\cite{Mamon05}, strong and weak gravitational lensing~\citep{Kochanek04}, and~the assumption of hydrostatic equilibrium, which requires knowledge of both the density and pressure (or temperature) profiles \citep{Arnaud01,Melin23}. To~recover masses from isolated observables like the thermal SZ effect \citep{Carlstrom02} and the X-ray luminosity of the ICM \citep{Vikhlinin06}, one must first calibrate scaling relations between these observables (e.g., $Y_{\rm SZ}$ and $L_{\rm X}$) and the total mass \citep{Lovisari20}.

For the kinematic approach, treating member galaxies as a relaxed system of collisionless tracers gravitationally bound within the cluster potential allows for mass estimation. Under~the isotropic assumption, the~total virial mass can be scaled globally via the virial theorem as $M_{\rm vir} \sim \sigma_v^2 R_{\rm vir} / G$, where $\sigma_v$ is the observable line-of-sight velocity dispersion~\cite{Binney08}. To~resolve the continuous radial mass profile $M(<r)$, one must solve the spherical Jeans equation for a steady-state system:
\begin{equation}
    M(<r) = -\frac{r \sigma_r^2}{G} \left( \frac{d\ln \nu}{d\ln r} + \frac{d\ln \sigma_r^2}{d\ln r} + 2\beta \right),
\end{equation}
where $\nu(r)$ is the spatial number density profile of the tracer galaxies, $\sigma_r(r)$ is their radial velocity dispersion, and~$\beta(r) \equiv 1 - \sigma_\theta^2/\sigma_r^2$ is the velocity anisotropy parameter characterizing orbital eccentricity~\cite{Carlberg96,Lokas01}.

The thermodynamic state of the diffuse intracluster gas provides an alternative dynamical weight. Assuming the ICM resides in hydrostatic equilibrium, its enclosed mass profile, $M(<r)$, can be derived from the thermal gas pressure profile, $P$, and~the gas density profile, $\rho_g$, via:
\begin{equation}
    \frac{1}{\rho_g} \frac{dP}{dr} = -\frac{GM(<r)}{r^2}.
\end{equation}

These techniques consistently reveal that the baryon budget comprises roughly $20\%$ of a cluster's total dynamical mass, unequivocally requiring a dominant dark component to sustain hydrostatic balance \citep{Pratt19}. This mass deficit is independently corroborated by gravitational lensing, which maps total mass distributions entirely free from dynamical state assumptions~\cite{Kneib11,Monteiro-Oliveira22b}. In the weak lensing regime, the observable convergence, $\kappa$, is directly proportional to the projected surface mass density:
\begin{equation}
    \kappa(\theta) = \frac{\Sigma(\theta)}{\Sigma_{\rm crit}}, \quad \Sigma_{\rm crit} = \frac{c^2}{4\pi G} \frac{D_s}{D_d D_{ds}},
\end{equation}
where $\Sigma_{\rm crit}$ is the critical surface mass density dependent on the angular diameter distances to the lens ($D_d$), to the source ($D_s$), and between the lens and source ($D_{ds}$) \cite{Umetsu20}. Extensive lensing surveys systematically reconstruct deep mass potentials that far exceed the visible baryonic content, confirming the spatial distribution of dark matter halos~\cite{Abbott22,Okabe25,Chiu25}.

\subsection{Numerical Cosmology: Hydrodynamical Simulations and Baryonic~Feedback}
\label{sec:numerical_alternatives}

The modern theoretical consensus surrounding structure formation is intrinsically bound to high-resolution cosmological simulations, which serve as irreplaceable computational laboratories for tracing the non-linear evolution of the Universe~\cite{Vogelsberger20}. By~initializing CDM density fields at high redshift (typically via second-order Lagrangian perturbation theory) gravity-only models mathematically demonstrate that CDM naturally aggregates into a complex topology of interconnected filaments, expansive voids, and~massive nodal halos. The~dark-matter-only baseline runs of contemporary flagship suites accurately evolve this hierarchical clustering, matching statistical observations of the large-scale cosmic web with remarkable precision~\cite{Frenk12}.\newpage

To explicitly bridge the physical gap between these invisible dark matter halos and luminous galaxies, state-of-the-art cosmological suites, such as IllustrisTNG~\cite{Springel18}, EAGLE~\cite{Schaye15}, Magneticum~\cite{Dolag16}, and~FLAMINGO~\cite{Schaye23}, couple this collisionless gravitational backbone with complex subgrid baryonic physics. These advanced full-physics models demonstrate that deep, pre-existing dark matter potential wells are a fundamental prerequisite for early gas cooling, star formation, and~the subsequent regulation of galactic growth via energetic feedback~\cite{Somerville15}. Without~a massive, collisionless dark component driving the initial gravitational collapse, hydrodynamical models cannot generate structures rapidly enough to match the cosmic timeline; consequently, they fail to reproduce the observed stellar mass functions and the morphological diversity of galaxies in the local Universe~\cite{Vogelsberger14}.

\section{Modified Gravity Versus Dark Matter: Evaluating the MOND~Alternative}
\label{sec:mond}

As established in the previous section, the~$\Lambda$CDM paradigm is observationally indispensable for explaining the macroscopic evolution of the Universe; a massive, collisionless scaffolding is an absolute prerequisite for structure formation, from~primordial CMB anisotropies to the large-scale cosmic web. However, despite its overwhelming success on these cosmological scales, the~persistent lack of direct laboratory detection of the dark matter particle has historically motivated phenomenological alternatives. These theoretical models attempt to circumvent the need for invisible mass by targeting the specific physical regime where dark matter was originally inferred: the anomalous, localized kinematics of individual galaxies. The~most prominent and enduring of these frameworks is Modified Newtonian Dynamics (MOND) \cite{Milgrom83a}.

Rather than adding dark matter, MOND proposes a fundamental modification of gravity in the weak-field regime \citep{Sanders02}. This is achieved by altering the Newtonian gravitational potential, $\Phi$, through a nonlinear Poisson equation:
\begin{equation}
    \nabla \cdot \left[ \mu\left(\frac{|\nabla\Phi|}{a_0}\right) \nabla\Phi \right] = 4\pi G\rho,
\end{equation}
where $\mu(x)$ is an interpolating function that transitions from $\mu(x) \approx x$ in the deep-MOND regime ($|\nabla\Phi| \ll a_0$) to $\mu(x) \approx 1$ in the Newtonian limit~\cite{Famaey12}. The~critical acceleration scale, empirically calibrated to $a_0 \approx 1.2 \times 10^{-10} \, \text{m/s}^2$, successfully aligns with the kinematics of diverse galactic~systems.

MOND is undeniably compelling on galactic scales. It naturally predicts the Radial Acceleration Relation (RAR) \cite{McGaugh16,Tam23} and the Baryonic Tully--Fisher relation with remarkably low scatter, accurately describing rotation curves and velocity dispersions in spiral galaxies and dwarf spheroidals without requiring dark matter halos~\cite{Lelli17}.

Despite these localized successes, MOND faces severe, often insurmountable challenges when extrapolated to \rmo {cluster and cosmological} scales~\cite{Dodelson06,Ettori19,Eckert22,Kelleher24}. Classical MOND cannot describe an expanding universe; therefore, relativistic extensions like the Tensor--Vector--Scalar (TeVeS) theory were developed to reconcile the framework with General Relativity~\cite{Bekenstein04}. However, even these advanced formulations systematically fail to simultaneously reproduce the precise acoustic peak morphology of the CMB and the linear matter power spectrum~\cite{Skordis06}. To~match {\it Planck} data, MOND frameworks are frequently forced to inject massive sterile neutrinos or extra relativistic degrees of freedom, effectively reintroducing unseen mass to save a theory explicitly designed to eliminate it~\cite{Angus09}.\newpage

While MOND provides an elegant kinematic description for isolated galaxies, its mathematical inability to orchestrate cosmological structure formation exposes a fatal flaw. This theoretical shortfall is ultimately crystallized by empirical observations. If~dark matter is merely a mathematical illusion caused by modified gravity, the~perceived gravitational lensing of a system must always strictly track its observable baryonic mass. To~definitively break this degeneracy, astrophysics requires an environment where the baryonic matter and the supposed dark matter have been physically ripped apart, a~scenario perfectly realized during massive galaxy cluster~collisions.

\section{Direct Evidence for Dark Matter: The Legacy of the Bullet~Cluster}
\label{sec:bullet_cluster}

Galaxy clusters are the most massive gravitationally bound structures in the Universe, assembling late in cosmic history (typically reaching virialization at redshifts $z \lesssim 1$) through the continuous hierarchical merging of smaller subclusters and groups \citep{Kravtsov12,Nelson24} (Figure~\ref{fig:halo_scaling}). These massive collisions are the most energetic physical events since the Big Bang, routinely dissipating kinetic energies on the order of $10^{63}$ to $10^{64}$ ergs over timescales of a few gigayears \citep{Sarazin04}. The~total mass budget of a fully virialized galaxy cluster is partitioned into three principal components, each governed by distinct dynamical regimes. The~stellar mass, confined mainly within the constituent galaxies, contributes only approximately 1--2\% to the total gravitational potential \citep{Gonzalez07}. The~dominant baryonic reservoir resides in the highly collisional ICM, a~diffuse, optically thin plasma that radiates primarily via thermal bremsstrahlung in the X-ray band---and accounts for roughly 10–15\% of the cluster mass \citep{Lagana13}. Ultimately, the~structural stability and global kinematics of the system are overwhelmingly dictated by the invisible, non-baryonic dark matter halo, which provides the gravitational binding for both the hot gas and the galaxies, constituting the remaining $\sim$85\% of the total mass budget \citep{Mulroy19}.

During a major cluster collision\endnote{Throughout this paper, we use the terms ``cluster merger'' and ``collision'' interchangeably.}, these three components are subjected to fundamentally different dynamical forces \citep{Feretti02}. Because~the mean free path of galaxies is orders of magnitude larger than the physical size of the cluster core, the~stellar components act, as~a first approximation, as~collisionless point masses, passing through one another with minimal impedance \citep{Clowe04}. In~stark contrast, the~diffuse ICM is governed by complex fluid dynamics \citep{ZuHone22}. As~the gas halos collide at relative velocities of several thousand kilometers per second, they experience extreme ram pressure, generating massive bow shocks, turbulent dissipation, and~violent deceleration \citep{Markevitch07}. Under~the $\Lambda$CDM paradigm, the~dominant dark matter halos are assumed to be strictly collisionless \citep{Blumenthal84}. Therefore, their phase-space evolution should mirror that of the galaxies, allowing them to pass through the hydrodynamical wreckage~unperturbed.

\begin{figure}[H]
\includegraphics[width=\textwidth]{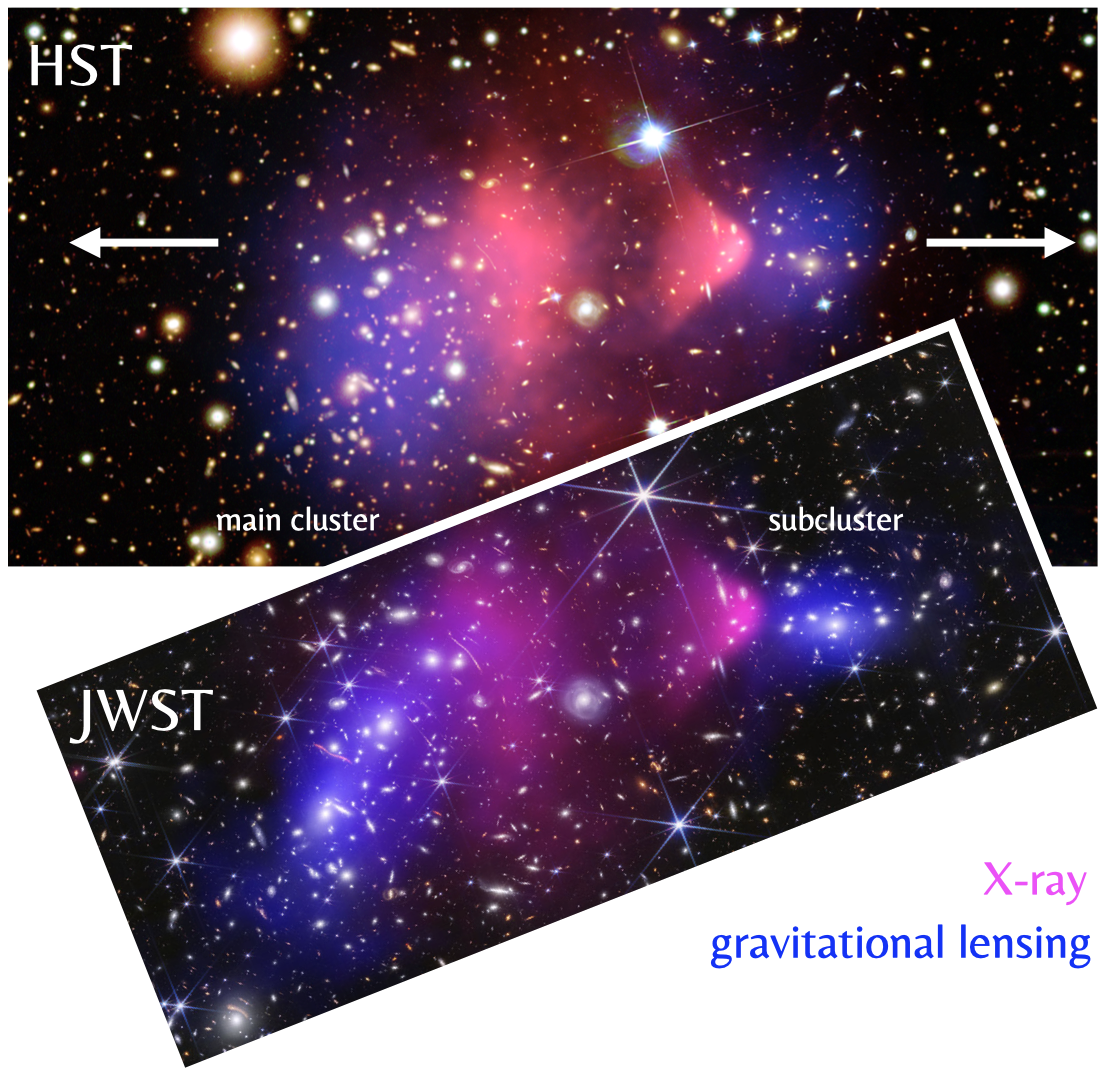}
\caption{Multi-wavelength composite of the Bullet Cluster (1E 0657-56). The~optical backgrounds are provided by the \textit{Hubble} Space Telescope (HST, top panel) and the \textit{James Webb} Space Telescope (JWST, bottom panel, registered to align with the HST spatial orientation). The~more massive main cluster is situated on the left, while the infalling subcluster is located on the right. The~ICM, mapped by \textit{Chandra} X-ray observations (pink), lags behind due to ram pressure, highlighting the prominent bow shock of the ongoing merger. Conversely, the~underlying total mass distribution, reconstructed via gravitational lensing (blue overlay), peaks at the collisionless galaxy concentrations rather than the collisional gas. The~white arrows indicate the trajectory of the interaction, as~the clusters move toward their respective apoapses. This macroscopic spatial separation between the baryonic gas and the mass centroids serves as direct empirical evidence for dark matter. 
 Image Credits: 
\textit{HST Component:} X-ray: NASA/CXC/CfA/M. Markevitch; Optical \& Lensing: NASA/STScI, Magellan/U.Arizona/D. Clowe, ESO WFI. 
\textit{JWST Component:} NASA, ESA, CSA, STScI, CXC; Science: J. Jee (Yonsei/UC Davis), S. Cha (Yonsei), K. Finner (IPAC/Caltech).}
\label{fig:bullet_composite}
\end{figure}

The Bullet Cluster (1E 0657--56, $z=0.296$) \citep{Clowe06} stands as the most historically significant example of this phase-space separation. Deep \textit{Chandra} observations reveal a prominent supersonic bow shock where the collisional ICM of the infalling ``bullet'' subcluster has been stripped and violently decelerated by the primary cluster's gas \citep{Markevitch02}. Conversely, weak and strong gravitational lensing reconstructions demonstrate that the system's dominant mass peaks have entirely bypassed this hydrodynamic drag, remaining strictly aligned with the collisionless galaxy distribution rather than the displaced gas \citep{Clowe06, Bradac06, Cho25}. The~dramatic spatial offset between the dominant baryonic reservoir and the underlying gravitational potential (see Figure~\ref{fig:bullet_composite}) confirms that the majority of the cluster's mass is effectively immune to hydrodynamic forces, behaving as a collisionless fluid on macroscopic~scales.

This failure of the gravitational potential to track the primary baryonic mass severely challenges MOND. Because~the vast majority of a cluster's baryonic mass resides in the ICM, MOND fundamentally predicts that the deepest gravitational potential (and thus the lensing mass peaks) should coincide with the X-ray gas \citep{Milgrom83b}. Instead, the~observed mass peaks trace the collisionless galaxies. Although~the complex spatial arrangement of colliding gas clouds can theoretically interact with MOND's non-linear equations to project a false gravitational peak away from the gas (creating ``phantom dark matter''), this geometric mirage is ultimately far too weak to account for the actual gravitational lensing observed \citep{Angus06}. Preserving the MOND framework therefore requires postulating a significant component of collisionless ``hot'' dark matter, such as massive neutrinos ($\sim$2~eV; see Figure~1 in \citet{Angus07}). More recently, high-resolution JWST observations have been used to propose a purely baryonic solution through the Integrated Galaxy-Wide Initial Mass Function (IGIMF; \citealt{Kroupa03}). Under~this framework, a~radically top-heavy star formation history in the cluster's early-type galaxies would leave behind massive, highly concentrated reservoirs of dark stellar remnants---a scenario ostensibly corroborated by the anomalously high metallicity of the X-ray gas \citep{Zhang26}. However, attempting to source the requisite lensing peaks from these unseen baryonic remnants merely trades one form of dark mass for another. Ultimately, these workarounds demonstrate that modified gravity cannot account for the system without fine-tuning large amounts of ``invisible'' mass, thereby lacking the general and self-consistent predictive power of the standard $\Lambda$CDM~paradigm.

\section{Tensions in the Collisionless Paradigm: The Small-Scale~Crisis}
\label{sec:small_scale_crisis}
\unskip

\subsection{Self-Interacting Dark Matter: Resolving Sub-Galactic~Anomalies}
\label{sec:sidm}

Despite the remarkable predictive success of the $\Lambda$CDM paradigm on macroscopic scales, persistent anomalies at sub-galactic scales continue to challenge our fundamental understanding of the dark sector \citep{Bullock17}. Chief among these ongoing tensions is the cusp-core problem: whereas dark-matter-only N-body simulations predict steep central density cusps (e.g., the~Navarro--Frenk--White profile \citep[NFW;][]{nfw96, nfw97}), kinematic observations of dwarf and low-surface-brightness galaxies frequently reveal shallower, constant-density cores \citep{Moore94}. Furthermore, the~field was long motivated by the \textit{missing satellites problem}, a~severe theoretical overprediction of low-mass subhalos compared to the observed census of Milky Way dwarf galaxies \citep{Klypin99}. However, this specific discrepancy is now widely understood to be largely resolved within the standard framework; deep sky surveys have uncovered a vast, previously hidden population of ultra-faint dwarf galaxies, while modern hydrodynamical simulations demonstrate that cosmic reionization and supernova feedback efficiently strip gas from the lowest-mass halos, rendering the majority of them completely dark \citep{Wetzel16, Simon19}. Yet, while baryonic mechanisms successfully mitigate the missing satellites issue, they fail to resolve the closely related \textit{``too big to fail''} problem. This kinematic discrepancy highlights that the most massive, dense subhalos predicted by simulations, those theoretically guaranteed to retain their gas and trigger star formation, are conspicuously absent or underdense in the local satellite population \citep{Boylan-Kolchin11}. The~standard framework also continues to struggle with the \textit{diversity problem}, the~unexpected scatter in the inner rotation curves of mass-matched galaxies \citep{Oman15}, and~the \textit{plane of satellites problem}, wherein satellite galaxies around the Milky Way and Andromeda exhibit highly correlated, planar alignments that are statistically rare in standard cosmological simulations \citep{Pawlowski12}.

To resolve these persistent sub-galactic anomalies while preserving the macroscopic triumphs of $\Lambda$CDM, Self-Interacting Dark Matter (SIDM) has emerged as one of the most compelling theoretical frameworks \citep{Spergel00, Tulin18}. This paradigm introduces a finite particle self-interaction cross-section per unit mass,
\begin{equation}
\sigma_m \equiv \frac{\sigma}{m} \quad [\mathrm{cm^2 \, g^{-1}}],
\label{eq:sigma_m_def}
\end{equation}
which characterizes the effective scattering probability during dynamic astrophysical interactions, ranging from localized halo thermalization to macroscopic cluster–cluster collisions \citep{Markevitch04, Randall08}. Typically parameterised within the range of $\sigma_m \sim 0.1$--$10 \text{ cm}^2\,\text{g}^{-1}$, this self-interaction fundamentally alters the phase-space evolution of dense, non-linear environments \citep{Kaplinghat16}. In~the high-density central regions of dark matter halos, these self-interactions facilitate efficient energy-momentum exchange \citep{Spergel00}. Because~the inner halo is dynamically colder than the velocity dispersion peak located further out, this scattering acts as an inward heat flux \citep{Kochanek00, Balberg02}. This thermalization process redistributes kinetic energy, causing the central particle orbits to expand \citep{Vogelsberger12}. Consequently, the~steep, collisionless density ``cusp'' predicted by CDM naturally flattens into an isothermal, constant-density ``core'' \citep{Rocha13,Robertson19}. By~physically lowering the central density and corresponding circular velocities of dwarf-scale halos, this core expansion elegantly and simultaneously resolves both the cusp-core and the ``too big to fail'' problems \citep{Vogelsberger12, Peter13}. Furthermore, when the gravitational potential of the baryonic distribution (such as a stellar disk) is coupled with this thermalized dark matter fluid, the~resulting core sizes become highly sensitive to the local baryon concentration \citep{Kaplinghat14}. This baryon-SIDM interplay naturally reproduces the wide scatter observed in the inner rotation curves of mass-matched galaxies, offering a robust theoretical solution to the diversity problem \citep{Kamada17, Creasey17}.

Crucially, because~this core-flattening mechanism is intrinsically driven by active heat transport, these expanded, isothermal configurations do not represent permanent, static endpoints. Instead, the~contemporary theoretical frontier of SIDM centers on the long-term, late-time thermodynamic evolution of these systems \citep{Adhikari25}, specifically, the~inevitable phenomenon of gravothermal core collapse \citep{TurnerH21}. Because~a SIDM halo behaves as a self-gravitating fluid with a negative heat capacity, its expanded core cannot maintain stability indefinitely \citep{Balberg02}. Over~cosmological timescales, the~core eventually becomes hotter than the surrounding outer halo, reversing the system's internal thermal gradient \citep{Koda11}. This inversion causes the inner halo to shed heat outward. Paradoxically, due to the negative heat capacity of gravitationally bound systems, this loss of thermal energy forces the core to contract under its own gravity and grow even hotter \citep{Pollack15}, accelerating the outward heat transfer and triggering a runaway thermodynamic instability \citep{Essig19}. Recent high-resolution cosmological simulations by \citet{Silverman26} reveal that this evolutionary path depends heavily on a halo's environmental history. While structurally quiescent halos rapidly succumb to gravothermal collapse, dynamically active mergers inject sufficient kinetic energy to stall the runaway process. Depending on these localized environmental conditions, the~late-time evolution of an SIDM halo can diverge dramatically: it can preserve an expanded core, compress it into an ultra-steep central cusp, or~even seed the direct collapse of supermassive black holes in isolated dwarf galaxies. Consequently, SIDM is no longer modeled simply as a static mechanism for core flattening, but~as a dynamic, time-dependent framework capable of producing the extreme structural diversity observed across the Universe \citep{Jiang23}.

\subsection{Bullet Cluster Constraints on~SIDM}
\label{sec:bullet_cluster_sidm}

Beyond its definitive role in establishing the physical existence of dark matter, the~Bullet Cluster serves as a preeminent macroscopic laboratory for placing direct empirical bounds on the SIDM cross-section \citep{Clowe06}. At~galactic scales, finite cross-sections slowly drive the core expansion detailed in Section~\ref{sec:sidm}. By~stark contrast, high-velocity cluster mergers probe SIDM under sudden, highly dynamic extremes. Within~these environments, collisional dark matter would produce distinct kinematic signatures, including spatial offsets between baryonic and dark components via ram-pressure-like drag, evaporative mass loss from the colliding subclusters, and~altered core geometries \citep{Tulin18}. To~translate these theoretical systematics into empirical limits, \citet{Markevitch04} formulated three independent~constraints:

\begin{itemize}
    \item Morphological drag: A highly collisional halo would behave as a fluid, experiencing ram-pressure drag and deceleration similar to the X-ray gas. The~scattering depth is $\tau_s = \sigma_m \Sigma_s$, where $\Sigma_s \simeq 0.2 \, \text{g cm}^{-2}$ is the subcluster's average mass surface density. The~presence of a significant spatial offset between the dark matter and the gas peaks implies that the dark matter did not experience this fluid-like deceleration, requiring $\tau_s < 1$. This yields a conservative upper limit of $\sigma_m < 5 \, \text{cm}^2\,\text{g}^{-1}$.

    \item Kinematic deceleration: The observed high velocity of the subcluster is in excellent agreement with its expected gravitational free-fall velocity, suggesting it has not lost significant momentum to collisional drag. The~accumulated velocity loss due to dark matter collisions is $\Delta v = (\overline{p}/m)\sigma_m \Sigma_m$, where $\Sigma_m \simeq 0.3 \, \text{g cm}^{-2}$ is the main cluster's mass column density along the trajectory, and~$\overline{p} \approx 0.1 \, m v_s$ is the average momentum lost by the subcluster per particle collision. By~conservatively requiring that the velocity loss compared to free-fall is less than $1000 \, \text{km s}^{-1}$, this limits the cross-section to $\sigma_m < 7 \, \text{cm}^2\,\text{g}^{-1}$.
    
    \item Evaporative stripping: The observed mass-to-light ratios of the subcluster ($M/L_B \simeq 280 \pm 90$, $M/L_I \simeq 170 \pm 50$) match those of relaxed main cluster systems, implying that the dark matter halo survived the core passage without shedding a significant fraction of its mass. This agreement restricts the allowed fractional mass loss to $f \approx 0.3$. During~the merger, elastic collisions can eject bound particles if the imparted speed exceeds the subcluster's internal escape velocity ($v_{esc} \simeq 1200 \, \text{km s}^{-1}$). The~theoretical fraction of particles lost is $\chi \tau_m = \sigma_m \Sigma_m [1 - 2(v'_{esc}/v_0)^2]$, where $v_0 \approx 4800 \, \text{km s}^{-1}$ is the collision velocity in the subcluster frame. Requiring that the predicted collisional mass loss does not exceed the observed limit ($\chi \tau_m < f = 0.3$) yields the most sensitive constraint of the study, $\sigma_m < 1 \, \text{cm}^2\,\text{g}^{-1}$, effectively excluding the cross-section intervals typically proposed to solve small-scale galactic anomalies \citep{Markevitch04}.
\end{itemize}

Subsequent studies have employed high-resolution N-body simulations, coupled with hydrodynamics, and~observationally motivated cosmological frameworks to further refine the Bullet Cluster constraints on the self-interaction cross-section, as~summarized in Table~\ref{tab:sidm_constraints}.

\begin{table}[H]
\caption{Constraints on SIDM cross-section ($\sigma_m$) from the Bullet~Cluster.\label{tab:sidm_constraints}}
	\begin{adjustwidth}{-\extralength}{0cm}
		\begin{tabularx}{\fulllength}{m{3.3cm}<{\centering}m{5cm}<{\centering}m{2cm}<{\centering}C}
			\toprule
			\textbf{Study} & \textbf{Method} & \textbf{\boldmath{$\sigma_m$} [cm$^2$/g]} & \textbf{Notes} \\
			\midrule
			Randall~et~al. \citep{Randall08} & Offsets, M/L$^*$, N-body sim & $<$1.25, $<$0.7 & Tighter limit with equal pre-merger M/L~assumption \\
			\midrule
			Rocha~et~al. \citep{Rocha13} & N-body + hydrodynamics & $\sim$0.1 & $\sigma_m = 1$ produces too large~cores \\
			\midrule
			Robertson~et~al. \citep{Robertson17} & N-body sim with SIDM & $\lesssim$2 & Methodology can affect~limits \\
			\midrule
			Banerjee~et~al. \citep{Banerjee20} & Cosmological sim, weak lensing & $\lesssim$2 & Observationally motivated~analysis \\
			\bottomrule
		\end{tabularx}
	\end{adjustwidth}
	\noindent{\footnotesize{* M/L refers to the mass-to-light ratio consistency test.}}
\end{table}

Most recently, high-resolution analyses leveraging \textit{JWST} have introduced a new layer of complexity to this picture, suggesting that the Bullet Cluster is a more intricate dark matter laboratory than previously assumed. Advanced strong and weak lensing reconstructions \citep{Cha25,Cho25,Rihtarsic26} resolve the ``main'' cluster not as a monolithic halo, but~as a structure comprising distinct subclumps. This granularity reveals that the system's dynamics cannot be fully captured by simple binary merger models. The~macroscopic separation between mass and gas remains robust; nevertheless, the~presence of such substructure implies that the interaction history of the dark matter halos is significantly more complex than the idealized scenarios traditionally used to derive scattering cross-section~limits.

The Bullet Cluster established a compelling proof-of-concept, demonstrating that galaxy clusters can serve as natural laboratories for probing the microscopic properties of dark matter. Its discovery, a~clear spatial segregation between dark matter and baryons during a cluster merger, provides an observational bridge between the large-scale structure of the universe and the fundamental physics of the dark sector. This result inspired a broader observational paradigm: if a single cluster collision can so effectively decouple dark matter from baryonic physics, then other galaxy clusters undergoing similar processes can be leveraged in much the same way. The~following section explores this generalized approach in~detail.

\section{Merging Galaxy Clusters as Dark Matter Particle~Colliders}
\label{sec:mergers_as_colliders}

Merging galaxy clusters act as macroscopic cosmic particle colliders, serving as a unique astrophysical probe of dark sector microphysics. The~critical diagnostic for SIDM lies in the trajectory of the dark matter halo relative to its constituent galaxies. Because~individual galaxies are separated by vast distances, the~stellar mass behaves as a strictly collisionless fluid, establishing a pure, ballistic baseline. If~dark matter experiences any non-gravitational interactions, such as self-scatterings mediated by an exotic dark sector particle, it will experience a macroscopic drag force. Consequently, the~dark matter halo will undergo a physical deceleration relative to the unperturbed galaxies. This resulting spatial offset serves as a direct, observable signature of dark sector~collisionality.

\begin{figure}[H]
\includegraphics[width=\textwidth]{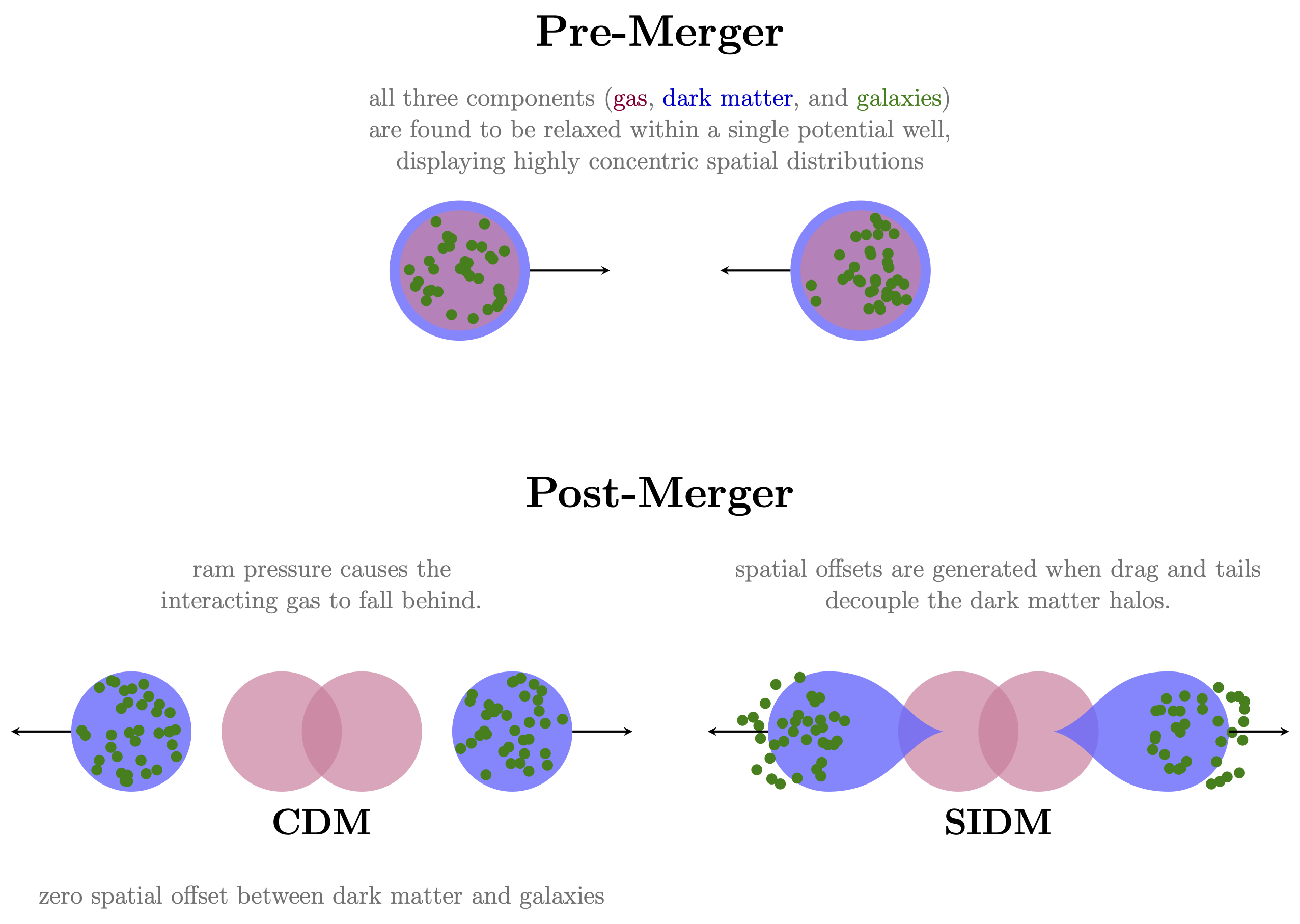}
    \caption{\textls[5]{Comparison of the distinct macroscopic observational imprints produced by CDM and SIDM following a cluster collision. \textbf{Top} (Pre-Merger): Prior to core passage, the~highly collisional gas, collisionless galaxies, and~dark matter are perfectly co-spatial within their respective gravitational potentials. \textbf{Bottom} (Post-Merger): After the collision, the~underlying microphysics of the dark sector dictate the spatial distribution of the mass components. Under~the standard CDM paradigm (\textbf{left}),}}
    \label{fig:cluster_collision_evolution}
\end{figure}

The physical mechanism driving this kinematic separation is illustrated in Figure~\ref{fig:cluster_collision_evolution}. In~the pre-merger phase, the~highly collisional intracluster gas, collisionless galaxies, and~dark matter are perfectly co-spatial, resting at the minimum of their shared gravitational potential. During~the violent core passage, the~purely collisionless galaxies traverse the merger interface completely unperturbed. Under~the standard CDM paradigm, dark matter is similarly collisionless and therefore perfectly traces this leading stellar distribution. Conversely, in~the SIDM framework, scatterings between the interpenetrating dark matter halos induce an effective macroscopic friction. This continuous momentum exchange causes the dark matter to deform, systematically decelerate, and~ultimately lag behind the uninhibited galaxies, generating a distinct structural offset in the post-merger~configuration.
\vspace{-15pt}

{\captionof*{figure}{dark matter is strictly collisionless and perfectly traces the ballistic, unperturbed trajectories of the galaxies, leaving only the ram-pressure-stripped gas lagging behind. Conversely, in~the SIDM framework (\textbf{right}), dark-sector self-scatterings induce a continuous macroscopic drag. This results in a unique, observable spatial segregation: while the purely collisionless galaxies lead the system, the~self-interacting dark matter halo systematically lags behind the stellar component, creating a distinct spatial offset between the dark mass and the~galaxies.}

\vspace{18pt}}

The subsequent sections establish the geometrical, statistical, and~computational frameworks deployed to parameterize these spatial offsets. Beyond~these analytical foundations, we synthesize ongoing observational efforts to reconstruct the intricate kinematics of these cosmic colliders, demonstrating how their macroscopic dynamics are translated into rigorous constraints on particle~physics.

\subsection{Statistical Ensembles of Cosmic Colliders: Merger Geometry and the Bulleticity~Framework}
\label{subsec:geometry}

Deriving strict three-dimensional constraints from two-dimensional astronomical images is an inherently degenerate problem. While the Bullet Cluster appears to bypass this issue due to an exceptionally rare viewing geometry, where the merger axis lies nearly perpendicular to our line of sight, the~vast majority of cosmic colliders unfold at arbitrary inclination angles. For~these typical mergers, the~projection from 3D space onto the 2D observational plane introduces significant systematic ambiguities, including unknown line-of-sight velocities, foreshortened physical separations, and~uncertain dynamical ages~\mbox{\citep{Dawson13, Harvey13}}. To~map these generic collisions to robust dark matter constraints, observers must therefore disentangle a series of interconnected geometric and kinematic~degeneracies.

\begin{figure}[H]
\includegraphics[width=\textwidth]{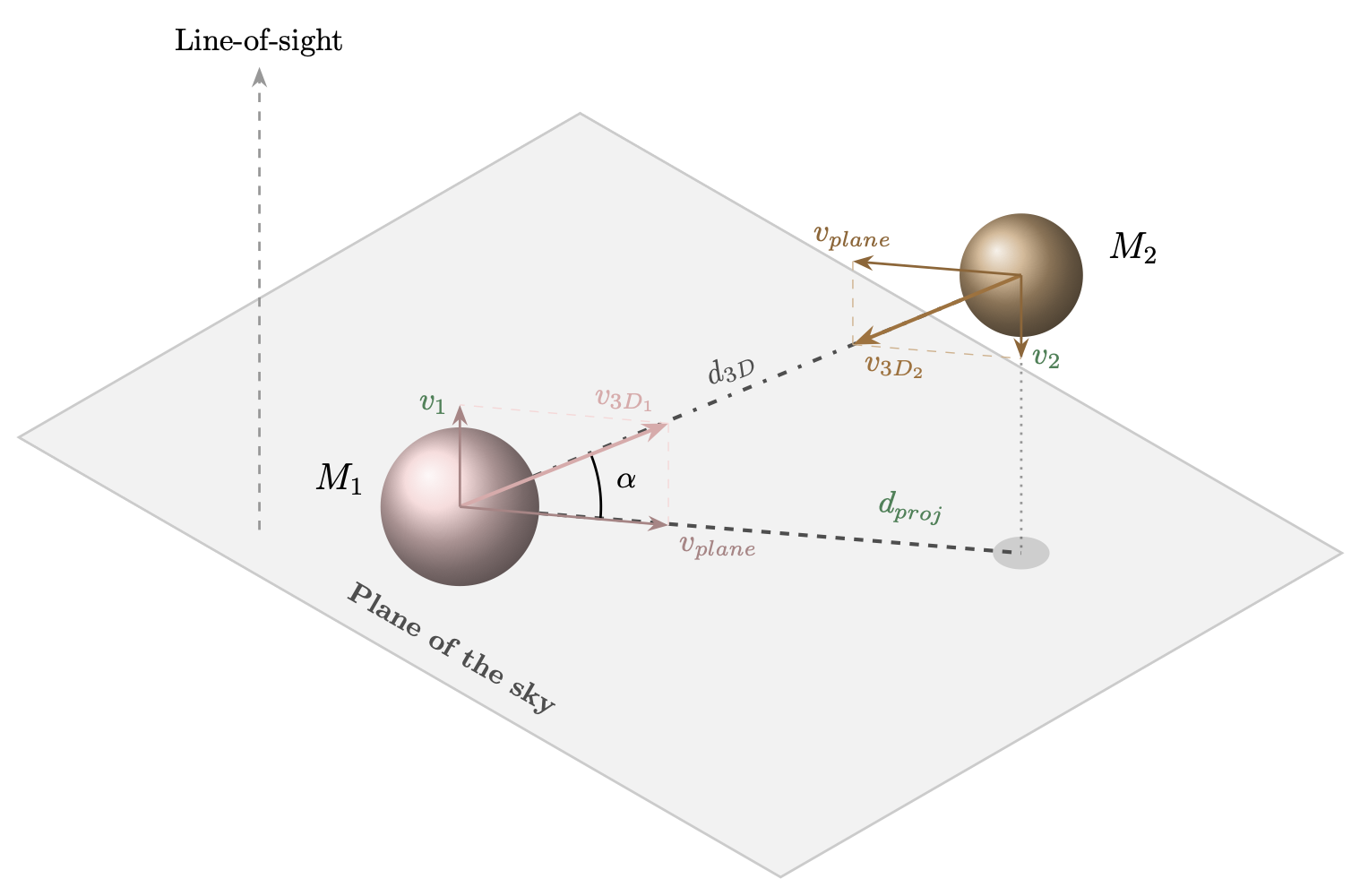}
    \caption{Geometric configuration of an idealized two-body cluster merger. The~schematic highlights the primary macroscopic properties of the system: the derived constituent halo masses ($M_1$ and $M_2$), alongside direct observables (highlighted in green) such as the line-of-sight radial velocities ($v_1, v_2$) and the two-dimensional projected separation on the sky ($d_{\text{proj}}$). Crucially, the~intrinsic three-dimensional counterparts ($d_{\rm 3D}$ and $v_{\rm 3D}$) cannot be directly accessed. This fundamental limitation arises because the inclination angle $\alpha$, which defines the orientation of the true merger axis relative to the plane of the sky, remains strictly unobservable, a~challenge further compounded by the fact that the transverse plane-of-sky velocity component ($v_{\rm plane}$) is entirely hidden for the vast majority of systems. Furthermore, although~this schematic depicts an outgoing system, the~merger phase itself constitutes an additional temporal degeneracy, as~single observational snapshots cannot inherently disentangle incoming from outgoing~configurations.}
    \label{fig:merger_geometry}
\end{figure}

The most immediate of these challenges stems from the unknown inclination angle $\alpha$ of the merger axis relative to the plane of the sky, which simultaneously drives both a \textit{projection degeneracy} and a \textit{kinematic degeneracy}. Spatially, this missing parameter restricts observations to sky-plane distances, meaning the true three-dimensional physical separation between dark matter and gas is severely foreshortened. Kinematically, the~same geometric ambiguity prevents the deprojection of spectroscopically measured line-of-sight radial velocities. While the transverse plane-of-sky velocity component ($v_{\rm plane}$) can occasionally be derived in rare cases via thermodynamic analysis---where distinct X-ray shock fronts provide density and temperature discontinuities that reveal the shock's Mach number and transverse collision speed \citep{Markevitch07,Springel07,Machado15b,Mello-Terencio}---such features are absent in the vast majority of systems. For~nearly all colliders, $v_{\rm plane}$ remains entirely unobservable, leaving the true three-dimensional collision velocity, and~thus the magnitude of the expected momentum transfer, highly uncertain. Both manifestations of this geometric ambiguity are illustrated in Figure~\ref{fig:merger_geometry}.

Even if the exact collision geometry could be fully determined, observers are still limited by a \textit{temporal (or merger phase) degeneracy}. Because~images capture only a single epoch, they freeze a dynamic process in an unknown frame, obscuring the time since pericenter passage. Without~this timeline, it is often impossible to determine whether subclusters are incoming or outgoing following a recent pericentric passage. This temporal degeneracy introduces a critical ambiguity: because the dark matter–ICM separation oscillates throughout the merger sequence \citep{Kim17, Fischer21}, a~small observed offset is difficult to interpret. Such a signal could arise from strong self-interaction drag if the system is caught at apocenter, when the displacement is momentarily reduced; alternatively, it could simply indicate an early-stage merger, where collisionless dark matter has not yet had sufficient time to separate from the gas. Both scenarios produce similarly small offsets, yet they correspond to radically different microphysical models~\citep{Dawson13}.

Beyond the timing of the collision, the~intrinsic orbital path introduces a significant \textit{impact parameter degeneracy}. Because~2D images cannot reveal the true 3D collision angle, researchers typically simplify dynamical models by assuming a strictly head-on collision with zero impact parameter \citep{Dawson13}. However, if~the system underwent a peripheral encounter, the~interacting halos would follow curved, non-linear orbits \citep{Doubrawa20}. When projected onto the observational plane, these trajectories can artificially segregate the mass components. Such projection effects can mimic the spatial lags expected from SIDM, or~they can mask them entirely---reinforcing why single-system merger analyses are insufficient for precision dark matter~constraints.

Untangling this web of geometric, temporal, and~orbital ambiguities within a single system requires relying on highly idealized assumptions; consequently, contemporary theoretical efforts have pivoted toward two complementary mitigation strategies. The~first approach leverages tailored computational simulations to map the phase space of specific individual systems, working backwards to reconstruct the merger timeline and constrain the kinematic parameters \citep{Mastropietro08, Bruggen12, Lourenco20, Chadayammuri21, Valdarnini24, Sirks24}. Recent methodological advances have introduced merger chronometers that can time-stamp the post-pericenter dynamical phase, helping to break the phase degeneracy. An~example is the shock-to-shock distance traced by double radio relics, which serves as a near-independent reference scale because the propagation speed of merger shocks is nearly insensitive to the self-interaction cross-section \citep{Jee26}. However, double radio relics are rare \citep{Lee26}, limiting the statistical power of the method. The~second strategy shifts the focus toward the statistical analysis of large cluster ensembles~\citep{Harvey13}. By~aggregating observations of numerous major mergers alongside continuous minor substructure infall events, this statistical framework naturally marginalizes over the random distribution of viewing geometries. Although~this ensemble methodology sacrifices the orbital history of any individual system, it successfully neutralizes the dominant projection biases, ultimately yielding potentially robust, geometry-independent constraints on the dark matter self-interaction cross-section \citep{Harvey14}. The~strength of these constraints, however, depends on the sample size and selection criteria, and~systematic biases can still affect the final limits if the ensemble is not carefully~constructed.

\begin{figure}[H]
\includegraphics[width=\textwidth]{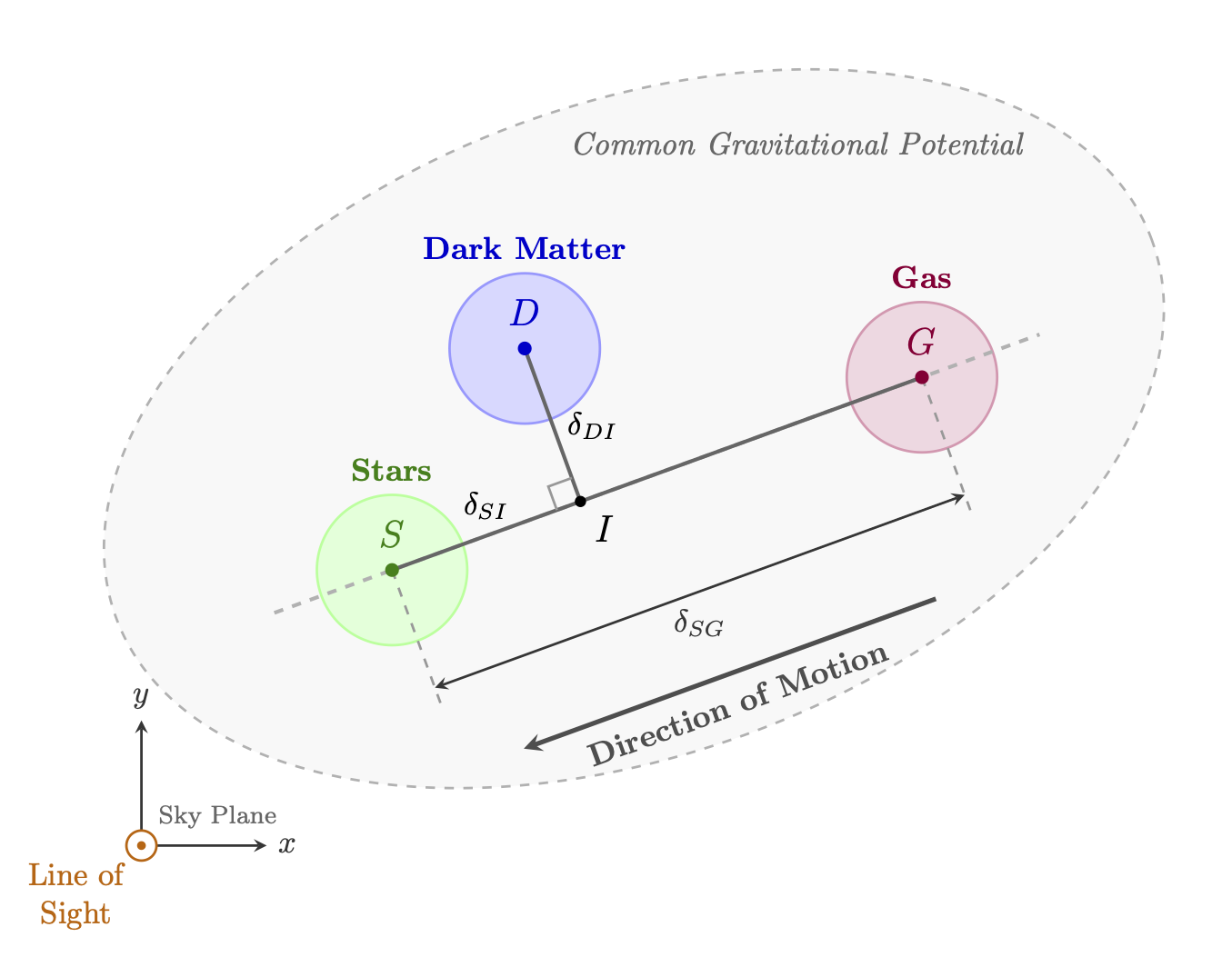}
    \caption{Schematic of the post-merger spatial configuration in a subhalo, viewed projected onto the sky plane ($x$-$y$). The~three primary mass components remain gravitationally bound within a common potential (dashed ellipse) but physically segregate due to varying interaction cross-sections. Operating as collisionless point masses, the~galaxies (Stars, $S$) lead the system. The~highly collisional gas ($G$) is severely decelerated by ram pressure, establishing the primary geometric axis and the total baryonic lag ($\delta_{SG}$). If~dark matter ($D$) undergoes self-interaction, it experiences an intermediate macroscopic drag. The~projection point $I$ decomposes this dark matter offset into a longitudinal displacement ($\delta_{SI}$) along the primary merger axis and an orthogonal displacement ($\delta_{DI}$).}
    \label{fig:post_merger_sidm}
\end{figure}

To implement this statistical ensemble approach, the~macroscopic separations shown in Figure~\ref{fig:post_merger_sidm} must be mathematically formalized. Each infalling substructure is modeled via its three principal mass components: the constituent galaxies ($S$), the~ICM gas ($G$), and~the dark matter halo ($D$). Acting as effectively collisionless point masses, the~member galaxies follow unperturbed ballistic trajectories. The~highly collisional ICM, by~contrast, experiences severe hydrodynamical ram pressure as it moves through the primary cluster potential, causing a visible deceleration. This predictable physical segregation, quantified by the projected spatial separation between the leading galactic centroid and the trailing gas ($\delta_{SG}$), empirically defines the substructure's forward axis of motion. If~the dark matter fluid possesses a non-zero self-interaction cross-section ($\sigma_m > 0$), macroscopic drag forces will cause its centroid to systematically lag behind the unperturbed stellar component, forming a total offset $\delta_{SD}$ between the dark matter and the stellar centroid. Geometrically decomposing this dark matter offset relative to the primary merger axis yields a longitudinal displacement ($\delta_{SI}$) and an orthogonal displacement ($\delta_{DI}$) \citep{Harvey14}.\vspace{-6pt}

Nevertheless, directly comparing these absolute projected offsets across a heterogeneous ensemble of cluster mergers introduces significant systematic biases, primarily driven by unknown line-of-sight projection effects. To~establish a universal geometric metric invariant to these observational degeneracies, the~physical displacements are formalized into a dimensionless fractional lag parameter, $\beta$, commonly referred to as \textit{bulleticity}~\citep{Massey11, Harvey14}:
\begin{equation}
    \beta \equiv \frac{\delta_{SI}}{\delta_{SG}}.
    \label{eq:bulleticity}
\end{equation}
This parametrization inherently neutralizes projection effects: because the 3D physical separations are foreshortened by the identical geometric projection factor ($\sin\alpha$, where $\alpha$ is the angle between the merger axis and the plane of the sky), their ratio perfectly cancels out the angular dependency, rendering $\beta$ an invariant reflection of the true structural geometry. Furthermore, decomposing the offset yields an orthogonal fractional component ($\beta_{\perp} \equiv \delta_{DI}/\delta_{SG}$) that serves as a critical, symmetry-driven null test. Assuming the underlying dark matter interactions are symmetric, a~statistically isotropic sample of mergers dictates that the ensemble mean $\langle \beta_{\perp} \rangle$ must be consistent with zero. Consequently, this orthogonal component provides a robust internal mechanism to test for systematic biases, where any measured deviation from zero directly quantifies the statistical uncertainty inherent in the positional estimates of the dark matter \citep{Harvey14, Harvey15}.

The dimensionless parameterization introduced above provides the mathematical scaffolding for a theoretical framework that translates the ensemble-averaged longitudinal lag, $\langle \beta \rangle$, directly into constraints on the SIDM cross-section, $\sigma_m$ \citep{Harvey14, Harvey15}. The~foundation of this analytical model rests on a specific microphysical hypothesis: dark matter self-interactions are frequent but exchange minimal momentum per collision. Such a regime is characteristic of scattering mediated by a light exotic particle, which generates long-range, highly anisotropic forces where momentum transfer is overwhelmingly parallel to the axis of motion. If~the underlying particle cross-section is velocity-independent \citep{Kahlhoefer14}, the~cumulative effect of these weak, frequent scatterings can be accurately modeled as a continuous, fluid-like macroscopic drag force that scales with the square of the subhalo's velocity ($\propto v^2$). By~mirroring the $v^2$ dependence of the hydrodynamical ram pressure acting on the gas, this framework allows the complex subhalo infall to be parameterized via a system of coupled differential equations. These equations govern the spatial segregation of the mass components by balancing the global gravitational pull of the primary cluster, the~component-specific drag forces, and~the local internal gravitational restoring forces---which effectively act as a temporary binding ``spring'' striving to keep the unperturbed galaxies, the~decelerating gas, and~the lagging dark matter unified during the collision~\citep{Harvey14}.

This system of coupled differential equations can be reduced to a simpler analytical form by recognizing that the macroscopic momentum transfer is governed by the dark matter halo's effective optical depth. As~the subhalo traverses the background ICM, gravitational restoring forces attempt to keep the gas, dark matter, and~galaxies aligned. However, the~extra drag force induces a fractional spatial lag that asymptotically approaches a maximum value dictated by how ``optically thick'' the halo is to self-interactions. This dynamical behavior is formalized by the analytical expression \citep{Harvey14, Harvey15}:
\begin{equation}
    \beta = B \left( 1 - \exp\left[-\frac{\sigma_m - \sigma_{\rm m, gal}}{\sigma^*_m}\right] \right)
    \label{eq:harvey_model}
\end{equation}
where adopting the notation of Equation~(\ref{eq:sigma_m_def}), $\sigma_{\rm m, gal}$ represents the interaction cross-section of the member galaxies (effectively zero for collisionless stellar systems), while the coefficient $B$ encodes the complex relative dynamical behavior between the dark matter and the gas, and~the denominator $\sigma^*_m$ represents the characteristic cross-section at which a halo of a given geometry becomes completely optically thick to~scattering.

To extract conservative limits from observational data, the~framework relies on two boundary assumptions that fix the free parameters introduced above: the stellar component is treated as strictly collisionless ($\sigma_{m, \rm gal} \approx 0$), and~the dark matter is assumed to experience the maximum possible drag relative to the gas ($B \approx 1$) \citep{Harvey15}. Additionally, the~analysis marginalizes over the characteristic optical depth by adopting $\sigma^*_m \approx 6.5 \pm 3~{\rm cm}^2\,{\rm g}^{-1}$ \citep{Harvey15}. Under~these parameterized conditions, any measured deviation of the ensemble's $\langle \beta \rangle$ from zero serves as a direct probe of the dark matter self-interaction cross-section. Ultimately, by~mathematically linking these macroscopic spatial offsets to the underlying particle kinematics, this analytical formalism establishes a robust bridge between large-scale cosmic structure and the subatomic physics of the dark~sector.

Building upon this analytical foundation, \citet{Harvey15} pioneered the observational application of this statistical ensemble framework by systematically analyzing a large catalog of colliding galaxy clusters, comprising 72 distinct dynamical substructures across 30 merging systems. To~independently isolate the three principal mass components, the~authors combined high-resolution optical imaging from the \textit{Hubble} Space Telescope with X-ray maps from the \textit{Chandra} X-ray Observatory. This multi-wavelength approach allowed them to map the underlying dark matter through weak gravitational lensing, pinpoint the collisionless stars via optical photometry, and~trace the highly collisional gas through its X-ray emission. The~authors found that the dark matter centroids remained tightly anchored to their host galaxies, yielding a mean spatial offset of just $5.8 \pm 8.2~{\rm kpc}$ and a corresponding longitudinal lag of $\langle \beta \rangle = -0.04 \pm 0.07$ (68\% confidence level). Translating this constraint through Equation~(\ref{eq:harvey_model}) yielded a highly restrictive upper bound on the self-interaction cross-section of $\sigma_m < 0.47~{\rm cm}^2\,{\rm g}^{-1}$ at the 95\% confidence~level.

While the tight constraints reported by \citet{Harvey15} represented an important methodological milestone, a~subsequent re-evaluation by \citet{Wittman17} demonstrated that these limits are subject to significant systematic uncertainties. Utilizing comprehensive multi-band imaging, deep spectroscopy, and~updated literature data to re-analyze the identical cluster ensemble, the~authors found that the original framework's strict reliance on single-band HST imaging introduced severe observational biases. Beyond~these data limitations, a~combination of instrumental and algorithmic vulnerabilities manifested in several distinct failures across the most heavily weighted substructures in the~ensemble. 

First, instrumental artifacts heavily corrupted the mass maps; for instance, in~the high-redshift cluster \textit{El Gordo} (ACTCL J0102-4915 \citep{Jee14}), \citet{Wittman17} revealed that a prominent dark matter peak mapped by \citet{Harvey15} was entirely spurious, probably originating from uncorrected data-cleaning issues directly within the telescope's CCD chip gap. Second, imperfect star-masking algorithms left residual light from exceptionally bright foreground stars misclassified as cluster galaxy luminosity centroids (e.g., in~MACS J2243.3-0935 \citep{vonderLinden14}), artificially shifting the apparent stellar positions by hundreds of kiloparsecs. Third, the~automated peak-matching routine used by \citet{Harvey15} frequently mistook low signal-to-noise ratio local X-ray maxima (induced by point sources or photon noise along smooth, extended gas ridges) for true physical gas peaks, yielding arbitrary galaxy–gas baseline vectors, as~is clearly seen in the subcluster A1758NE \citep{Monteiro-Oliveira17a}. Crucially, because~the inverse-variance weighting framework caused the statistical ensemble's signal to be dominated by only a few highly weighted substructures rather than being evenly distributed, these large localized mismeasurements failed to cancel out over the sample of 72 substructures, severely skewing the final cross-section~constraint.

Beyond these mapping challenges, the~re-evaluation highlighted key theoretical limitations within the underlying analytical model itself. The~optical depth prescription (Equation~\eqref{eq:harvey_model}) relies on a simplified scattering framework that assumes relatively small spatial displacements ($\le 30~{\rm kpc}$); however, \citet{Wittman17} demonstrated that 67 out of the 72 analyzed substructures departed from this regime, pushing the framework into a configuration where the assumed analogy between stellar and gaseous restoring forces breaks down. Furthermore, the~model implicitly treats the displacement ratio ($\beta$, Equation~(\ref{eq:bulleticity})) as a temporal constant, omitting the complex dynamics that occur after pericenter passage, when the effective drag force subsides and the components begin to dynamically fall back or slosh through the dark matter halo. Because~this post-pericenter behavior can alter or even invert the physical sign of the spatial offset over time, averaging the ensemble without accounting for individual merger phases introduces a notable systemic bias. Ultimately, \citet{Wittman17} concluded that accounting for these limitations significantly weakens the previously inferred statistical constraint, relaxing the SIDM cross-section limit to a more conservative upper bound of $\sigma_m \le 2~{\rm cm}^2\,{\rm g}^{-1}$ at the 95\% confidence~level.

\subsection{Numerical Frontiers: Simulating the Complex Physics of Cosmic~Colliders}
\label{subsec:numerical_frontiers}

Although empirical studies of merging galaxy clusters offer profound insights into the dark sector, their explanatory power is fundamentally limited by a static perspective. Telescopes capture only a singular, two-dimensional projection of highly non-linear cosmic collisions that unfold over gigayear time-scales \citep{ZuHone16}. Overcoming this temporal bottleneck requires state-of-the-art $N$-body and fully coupled cosmological hydrodynamical simulations, which provide a dynamic framework for the self-consistent forward-modeling of mergers under precisely controlled initial conditions \citep{Springel07,Molnar13,Molnar18,Molnar20,Zhang20,Valdarnini24}. By~acting as an indispensable bridge between microscopic particle physics and macroscopic astrophysical phenomena, these advanced simulations establish the robust theoretical baseline necessary to accurately decode the fleeting, complex snapshots captured on the sky \citep{Machado13,Machado15a,Machado15b,Doubrawa20,Machado24,Perroni26}.

Accurately measuring spatial offsets from single-epoch snapshots requires a precise definition of the centroid of each mass component. This is not a trivial matter: SIDM mergers\endnote{This statement holds true for major mergers as well, irrespective of the underlying mass ratio.} produce highly asymmetric spatial distributions, with~a dense core trailed by a diffuse, non-Gaussian tail of decelerated dark matter (see Figure~\ref{fig:cluster_collision_evolution}) \citep{Kim17}. For~a cross-section of $\sigma_m = 1~{\rm cm}^2\,{\rm g}^{-1}$ \citep{Markevitch04}, the~true separation between the galaxies and this core is remarkably small, typically $\le$$20~{\rm kpc}$ \citep{Kim17}. Consequently, the~measured offset ($\delta_{SI}$) is overwhelmingly dictated by the choice of analysis pipeline \citep{Robertson17}. Small smoothing kernels or parametric weak lensing shear fits successfully isolate the asymmetric core, yielding expected offsets of just $\sim$$10~{\rm kpc}$, a~signal entirely obscured by the $\pm 40\text{--}60~{\rm kpc}$ inherent lensing uncertainty \citep{Harvey15}. Conversely, larger smoothing scales or geometric algorithms like shrinking circles blend the core with the trailing wake, artificially shifting the centroid backward and inflating $\delta_{SI}$ to $\sim$$40~{\rm kpc}$ \citep{Kim17, Robertson17}.

Beyond these algorithmic constraints, characterizing the dark sector is heavily impeded by systematic observational noise that introduces false positives in SIDM searches. While large-scale cosmological hydrodynamical simulations confirm that the intrinsic, physical offset between the brightest cluster galaxy (BCG; \citep{Zenteno25}) and the dark matter centroid in a $\Lambda$CDM paradigm is sub-kiloparsec ($\sim$$1~{\rm kpc}$), standard observational tracers routinely introduce a false ``noise floor'' of $\sim$$10\text{--}20~{\rm kpc}$ \citep{Roche24}. Due to hydrodynamical non-equilibria and line-of-sight projection effects, utilizing gas centroids or strong-lensing emulations systematically inflates the measured separation by factors of 10 to 30, producing offsets that are observationally indistinguishable from genuine SIDM signatures \citep{Roche24}. This exceptionally low signal-to-noise ratio directly motivates the shift toward coordinate-independent metrics such as the bulleticity $\beta$ (Equation~\eqref{eq:bulleticity}), which normalizes the dark matter lag relative to the ram-pressure-stripped gas to minimize viewing-angle variances \citep{Sirks24}. As~an illustration, for~a cross-section of $\sigma_m = 0.1~{\rm cm}^2\,{\rm g}^{-1}$, the~true physical SIDM lag accounts for a mere $\sim$$5\%$ of the total gas displacement \citep{Sirks24}. Because~this subtle collisional signal is completely subsumed by the systematic artifacts of standard $\Lambda$CDM observations, individual cluster configurations might be structurally insufficient to isolate self-interactions. Disentangling a genuine collisional drag ($\beta > 0$) from standard astrophysical variance requires expanding beyond hand-picked, highly visible mergers to analyze large, unbiased statistical samples of $\sim$100 merging systems \citep{Sirks24}---a highly feasible number for the next generation of surveys like LSST \citep{Ivezic19}, Euclid \citep{Laureijs11}, and~the Nancy Grace Roman Space Telescope \citep{Spergel15}.

To disentangle genuine SIDM signatures from astrophysical variance, high-fidelity $N$-body and fully coupled hydrodynamical simulations serve as an essential complement to direct observations. Unlike idealized analytical models that treat galaxies as bare, collisionless test particles, realistic simulations must account for galaxies remaining deeply embedded within their own dark matter subhalos during collisions \citep{Kummer18}. Self-interactions with the ambient intra-cluster dark matter induce both a decelerating drag force and severe evaporation (mass loss) on these subhalos. Because~the galaxy is gravitationally anchored to this decelerating host, its trajectory is altered, preventing it from decoupling as a purely collisionless tracer. Consequently, the~true physical separation between the constituent galaxies and the dark matter centroid is substantially diminished. Frameworks that treat galaxies merely as bare test particles can systematically overestimate expected macroscopic offsets by up to $\sim$5~kpc, demonstrating that the internal structure of subhalos must be explicitly incorporated to accurately extract SIDM cross-section limits (see Figure~5 in~\citep{Kummer18}).

Characterizing the dark sector through a single, integrated cross-section $\sigma_m$ is an oversimplification that might mask complexities in the underlying particle physics. To~accurately map the macroscopic astrophysical observables generated within these cluster-scale colliders back to fundamental particle theories, one must refine the description of particle behavior, resolving, for~example, whether microscopic scatterings are dominated by rare, large-angle transfers or frequent, small-angle deflections. Physical observables diverge profoundly between models featuring rare, large-angle scatterings (rSIDM) and those characterized by frequent, small-angle scatterings (fSIDM) \citep{Kahlhoefer14}. In~both regimes, the~majority of the matter remains bound to the same gravitational potential, meaning the primary dark matter density peak always remains securely tied to the peak of the collisionless galaxies~\citep{Kahlhoefer14}. However, their spatial centroids diverge morphologically: in the rSIDM regime, rare hard scatterings eject a fraction of particles to form a backward-trailing dark matter tail, shifting the halo's overall spatial centroid backward. Conversely, in~the fSIDM regime, frequent interactions exert an effective macroscopic drag force that collectively decelerates the dark matter halo, causing a fraction of the unaffected galaxies to outpace the dark matter and form a forward tail, which shifts the galaxy centroid ahead \citep{Kahlhoefer14}.

Attempting to track this divergent spectrum of infinitesimal momentum transfers using standard discrete collision algorithms renders large-volume simulations computationally intractable. To~overcome this, hybrid numerical schemes (hSIDM) have been developed to unify the effective drag of small-angle scatterings with the discrete Monte Carlo treatment of large-angle scatterings \citep{Arido25}. Such implementations mathematically demonstrate that highly anisotropic scattering generates significantly larger spatial offsets than purely isotropic models \citep{Arido25}. When applied to full cosmological volumes, it becomes evident that while rSIDM and fSIDM produce largely similar macroscopic halo shapes---with fSIDM tending to generate slightly more spherical central geometries---frequent interactions (fSIDM) drastically suppress the abundance of satellite subhalos compared to rare interactions \citep{Fischer22}. Compounding the complexity of angular dependencies is the assumption of collision kinematics. Historical constraints have largely relied on the simplifying premise of a velocity-independent scattering rate; however, physically motivated particle models (e.g., light mediators) inherently predict a velocity-dependent cross-section \citep{Sabarish24}. Explicitly simulating velocity-dependent self-interactions across both rare and frequent scattering regimes demonstrates that the macroscopic dynamics of the merger are fundamentally altered by the relative velocities of the interacting halos \citep{Sabarish24}.

While isolating macroscopic dark matter-galaxy offsets remains a primary observational goal, these signatures are highly transient. In~standard velocity-independent models, offsets typically reach their maximum shortly after the first pericenter passage~\citep{Kahlhoefer14}. However, introducing a velocity-dependent cross-section drastically alters this temporal evolution; because high initial collision velocities at pericenter naturally suppress the scattering rate, the~spatial offset of the BCG experiences its most significant growth during the late, low-velocity phases of the merger \citep{Sabarish24}. By~capturing this late-stage offset growth, velocity-dependent simulations provide updated, more accurate upper limits on the SIDM cross-section, replacing previous boundaries derived from isotropic, velocity-independent approximations \citep{Sabarish24}. Alongside these velocity effects, the~mass ratio of the merging clusters fundamentally dictates offset survivability. While idealized simulations often focus on symmetric, equal-mass mergers because they maximize theoretical spatial offsets~\citep{Kim17}, such events are cosmologically rare. Consequently, realistic models must rely on unequal-mass collisions, where tidal forces and fSIDM-induced evaporation drive rapid subhalo dissolution \citep{Fischer21}. This extreme transience severely complicates the traditional approach of tracking surviving dark matter subhalos at late stages, necessitating a shift in observational strategy. Rather than hunting for offsets generated during the initial, high-speed core crossing, analyses should target late merger phases (between the first apocenter and second pericenter) where non-linear evolution amplifies the differences between interaction models \citep{Fischer21}. During~these advanced stages, the~faster dissolution of fSIDM subhalos leaves a distinct morphological signature: systems can evolve into states where the dark matter has fully coalesced into a single mass, yet the collisionless galaxies maintain separate, identifiable spatial concentrations \citep{Fischer21}. Searching for these distinct galactic components within a coalesced dark matter halo provides a powerful late-time metric to break degeneracies between rare and frequent scattering models \citep{Fischer21}.

The DM-only baseline provides an indispensable foundation, but~relying solely on dark matter physics to interpret late-time signatures remains physically incomplete. Real cluster mergers are permeated by a dense ICM that fundamentally shapes the system's overall dynamical evolution. Fully coupled hydrodynamical simulations demonstrate that this gaseous component acts as a powerful delayed amplifier for spatial offsets in SIDM mergers~\citep{Fischer23}. Specifically, although~gas physics has a negligible impact on dark matter–galaxy separations immediately after the initial core crossing, it dramatically boosts these offsets during the later apocenter phases \citep{Fischer23}. This baryonic amplification further cements highly evolved mergers as the premier observational targets for probing SIDM microphysics. Crucially, it renders macroscopic offsets detectable even for exceedingly small self-interaction cross-sections that would otherwise evade detection via standard relaxed cluster core size measurements \citep{Fischer23}. Beyond~standard hydrodynamics, recent numerical frameworks have begun exploring even more complex scenarios, including algorithms designed to simulate direct, non-gravitational dark matter–baryon scattering~\citep{Fischer25}. Nevertheless, the~primary observational challenge remains mapping the dominant variance introduced by the ICM. Consequently, robust limits on SIDM collisionality can no longer rely on isolated analytical models; rather, they must be statistically calibrated against the highly coupled astrophysical feedback predicted by full multiphysics simulations. Ultimately, it is only through these comprehensive numerical frameworks that we can confidently decode the static, complex snapshots captured by next-generation surveys into fundamental particle~physics.

\section{Challenges in Cluster Merger~Analyses}
\label{sec:challenges}
\unskip

\subsection{Asymmetric Dissociation: The Demi-Bullet~Morphology}
\label{sec:demi_bullet}

\begin{figure}[H]
    \subfloat[\centering]{
        \includegraphics[width=0.8\textwidth]{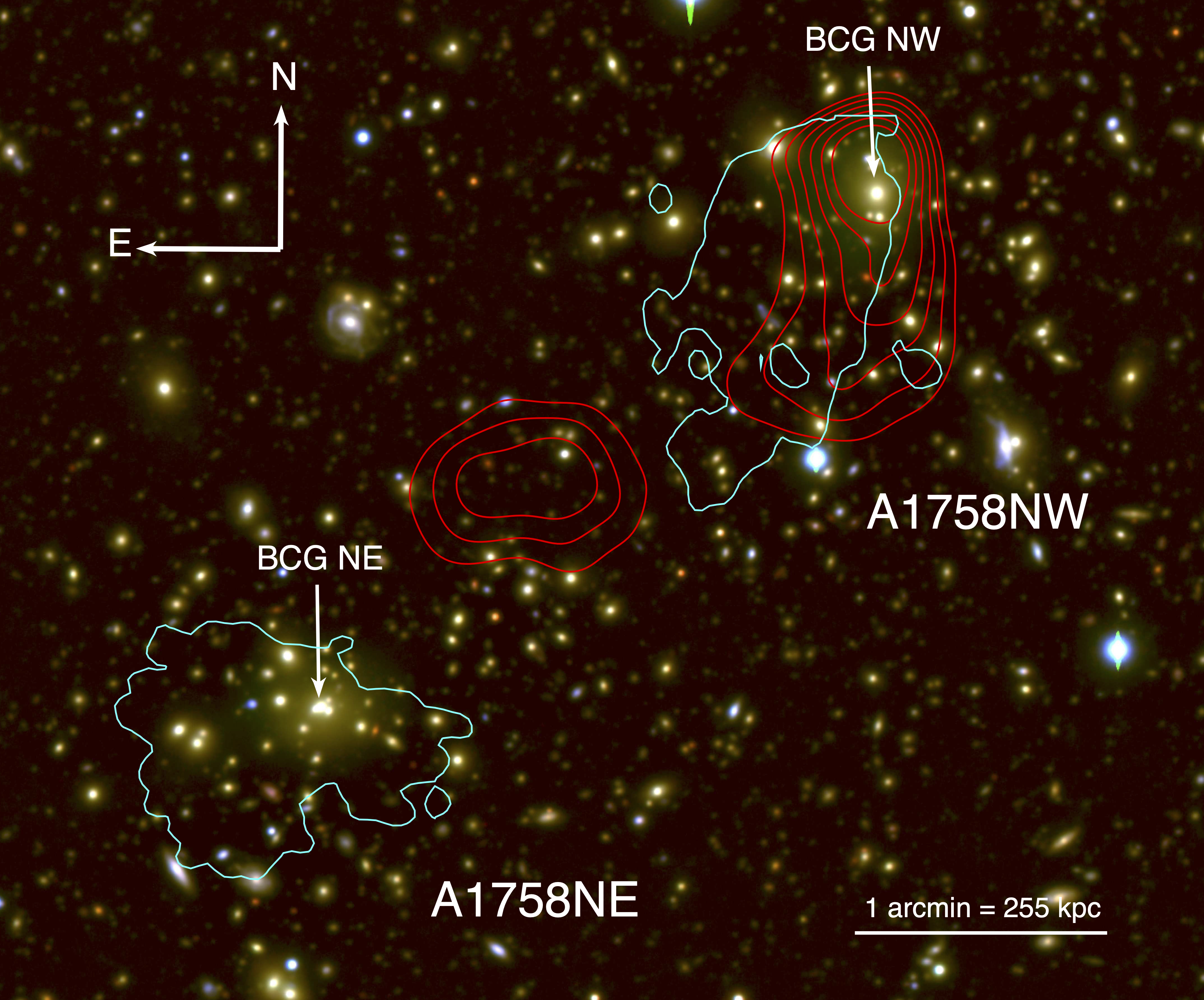}
        \label{fig:A1758}
    }
    \hfill 
    \subfloat[\centering]{
        \includegraphics[width=0.8\textwidth]{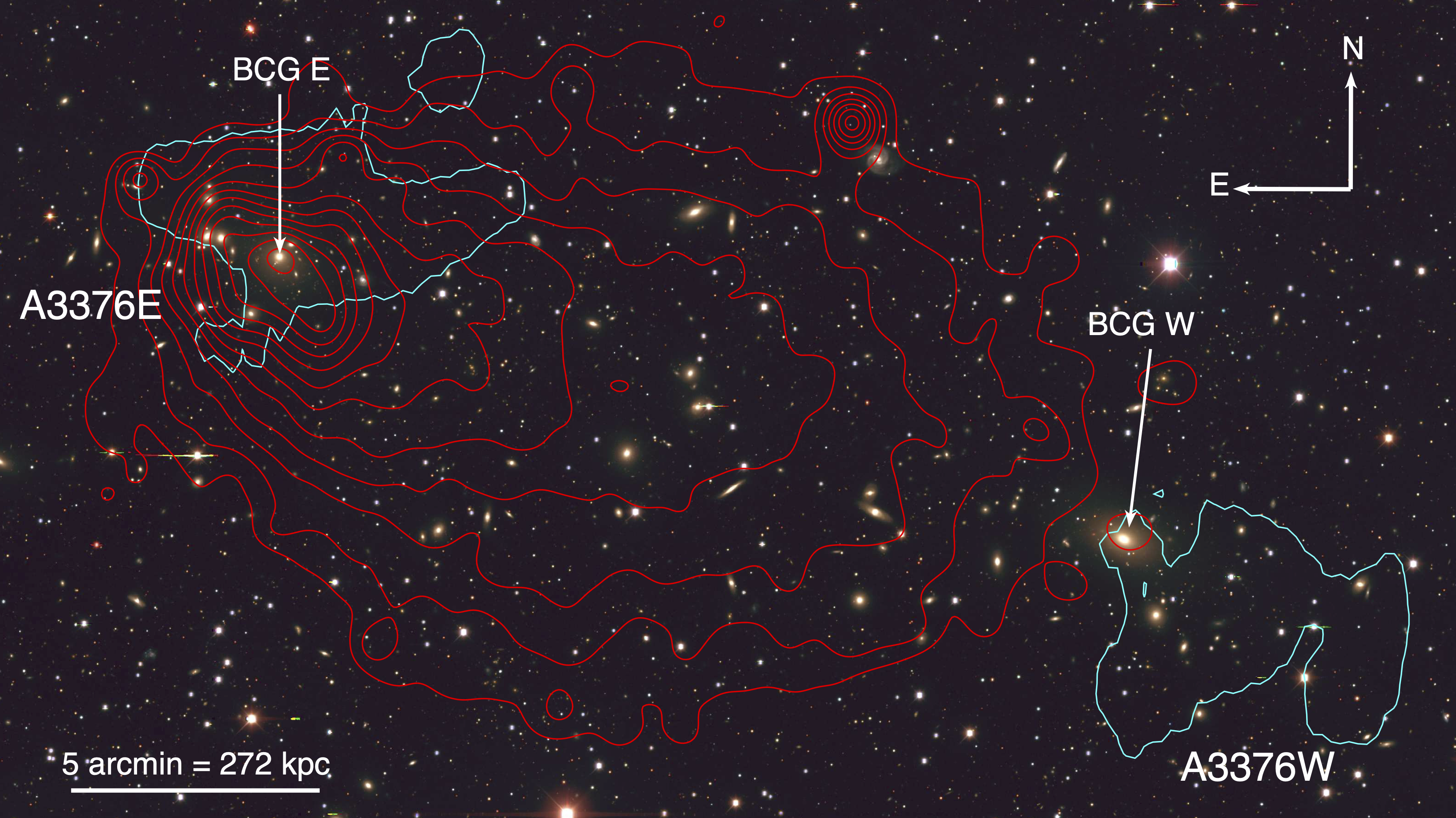}
        \label{fig:A3376}
    }
    \caption{Observational examples of demi-dissociative (demi-bullet) configurations in merging galaxy clusters. 
    (\textbf{a}) Abell 1758N: Combined optical $B R_C z^\prime$-band image with the ICM distribution traced by \textit{Chandra} X-ray emission (red). Cyan lines denote the $2\sigma$ confidence contours of the total mass center positions, determined through weak gravitational lensing, for~the northwest (A1758NW, right) and northeast (A1758NE, left) subclusters, respectively. The~NE subcluster demonstrates a classic demi-bullet decoupling, displaying a significant spatial separation ($96^{+14}_{-15}$~arcsec) between its X-ray gas peak and total mass center, while its corresponding BCG remains spatially coincident within $2\sigma$. Conversely, the~NW subcluster exhibits complete structural alignment among all three components within $2\sigma$ \citep{Monteiro-Oliveira17a}.
    (\textbf{b}) Abell 3376: Optical $g^\prime r^\prime z^\prime$-band overlay featuring \textit{XMM-Newton} X-ray contours (red) along with the $95\%$ (cyan) confidence level contours of the mass center positions, via weak gravitational lensing. The~X-ray gas distribution is starkly unimodal, with~the lone core peak associated exclusively with the eastern subcluster (A3376E), showing spatial alignment with both its mass center and BCG within the $95\%$ confidence level. Conversely, the~western subcluster (A3376W) showcases an extreme post-merger anatomy where the dark matter halo has been entirely stripped of any measurable gas counterpart, while its BCG tracks the lensing mass center \citep{Monteiro-Oliveira17b}.}
    \label{fig:demi_bullet}
\end{figure}

The Bullet Cluster is widely championed as the archetypal dissociative merger; however, statistically speaking, it represents a morphological exception. Beyond~the specific kinematic features detailed in Section~\ref{sec:bullet_cluster}, it is a rare ``double dissociative'' system, wherein both the primary and secondary subclusters exhibit distinct spatial offsets between their dark matter halos and collisional gas cores (as illustrated in Figure~\ref{fig:cluster_collision_evolution}). In~reality, the~vast majority of observed binary mergers manifest as ``demi-bullets'' (or demi-dissociative systems), in~which only a single substructure presents a measurable mass-gas offset \citep{Monteiro-Oliveira17a, Monteiro-Oliveira17b, Monteiro-Oliveira18, Monteiro-Oliveira20, Kelkar20, Monteiro-Oliveira21, Monteiro-Oliveira22a, Hernandez-Lang22, Pandge19}. Observationally, the~most common post-merger anatomy consists of a system in which at least one of the interacting subclusters successfully retains its collisional gas core---as exemplified by the merging system A1758N ($z\sim0.278$) \citep{Machado15b,Monteiro-Oliveira17a}, illustrated in Figure~\ref{fig:demi_bullet}a. In~more extreme scenarios, there is no measurable gas peak remaining associated with one or more of the dark matter halos, a~configuration clearly observed in the merging cluster A3376 ($z\sim0.046$; Figure~\ref{fig:demi_bullet}b).

Numerical models suggest that the physical mechanism driving this asymmetric or partial dissociation is deeply rooted in the complex interplay between hydrodynamic ram pressure and the initial structural parameters of the merging halos. Within~the standard $\Lambda$CDM framework, dedicated smoothed-particle hydrodynamics $N$-body simulations demonstrate that demi-dissociative morphologies naturally emerge under specific orbital configurations. For~example, models of A1758N evaluated roughly $0.4$~Gyr after first pericentric passage reveal that localized ram pressure efficiencies and mass ratios dictate the extent of gas stripping; the primary core retains sufficient gravitational binding energy to hold its gas intact, while the secondary component undergoes thorough stripping \citep{Machado15b}. Similarly, systematic modeling of the dissociative system Abell~2034 indicates that the magnitude of the resulting dark matter--gas offset is heavily influenced by the ratio of the interacting clusters' central gas densities, a~parameter that proves more critical than their dark matter density ratio \citep{Moura21}. Ultimately, these numerical findings confirm that the morphological diversity of cosmic colliders---ranging from symmetric double dissociation to asymmetric demi-bullets---does not unambiguously signal anomalous dark matter self-interactions. Instead, these structures can be fully accommodated within standard astrophysics as a natural consequence of localized boundary conditions, such as impact parameters, collision timelines, and~initial gas~concentrations.

\subsection{The Weak Lensing Mass Bias in Non-Equilibrium~Systems}
\label{sec:wlmb}

A critical challenge in accurately constraining the post-merger anatomy of galaxy clusters lies in the reliability of foundational mass measurements. Traditionally, the~total mass distribution of these cosmic colliders is reconstructed utilizing weak gravitational lensing techniques. However, these reconstructions systematically rely on fitting observational data to analytical density profiles (such as the NFW \citep{wright00} profile) which are mathematically derived under the assumption that the dark matter halos exist in a state of dynamical equilibrium \citep{Soja18}. During~energetic cluster collisions, particularly in epochs closely following the first pericentric passage, the~interacting dark matter halos undergo profound internal structural perturbations, severely violating the equilibrium assumption. Theoretical investigations indicate that forcing regular analytical profiles onto these heavily disturbed systems introduces a significant systematic error, designated as the weak lensing mass bias (WLMB). In~scenarios involving massive mergers ($M \sim 10^{15} M_\odot$), this bias has been shown to artificially overestimate the mass of the interacting subclusters by up to 80\% \citep{Lee23}.

The propagation of the WLMB raises fundamental concerns regarding the trustworthiness of follow-up tailored hydrodynamical simulations. Unlike broader cosmological simulations that evolve from early-universe initial conditions, tailored simulations of specific merging clusters are highly deterministic, relying directly on the observationally reconstructed mass fields to establish their initial boundary conditions. Consequently, if~the input lensing masses are intrinsically biased by the complex dynamic state of the cluster, the~subsequently derived orbital parameters and the estimated merger age (the time elapsed since the initial pericentric passage) may be significantly skewed. While some analyses of late major mergers suggest that the resulting discrepancy in the derived merger age might be relatively modest (on the order of $\sim$0.1~Gyr) \citep{Albuquerque24}, the~broader influence of biased initial conditions on the overarching merger history remains a source of critical~uncertainty.

The question of how thoroughly these simulated post-merger anatomies can be trusted remains definitively open. The~magnitude of the WLMB is not a static constant; its impact is highly degenerate, depending intricately on both the observer's specific line-of-sight and the time-dependent evolution of the collision itself. Clarifying these uncertainties requires the generation of extensive mock observational datasets and systematic comparative simulations. Such ongoing efforts are essential to either validate the historical parameters derived from past hydrodynamical models or to establish robust methodologies capable of mitigating this mass bias in future observational~campaigns.

\subsection{Offset Comparisons with Simulations: Observational and Numerical~Challenges}
\label{sec:offset_challenges}

The interpretation of merging galaxy clusters as precision laboratories for SIDM hinges on the accurate measurement of projected dark matter--galaxy offsets, $\delta_{\rm SD}$ \citep{Harvey15,Wittman17}. Reconstructing the post-collision anatomy requires the joint interpretation of multi-wavelength observables, each with distinct systematic uncertainties including centroiding ambiguities, finite angular resolution, and~line-of-sight contamination \citep{Markevitch07,Dawson13,Simet17,Ding25}. Consequently, the~inferred offsets depend on centroid definitions and lensing mass reconstruction techniques \citep{Harvey16,Lee23}. In~addition, the~inherent two-dimensional projection of a transient event prevents the unique determination of merger inclination, orbital geometry, and~evolutionary phase \citep{Markevitch07,Wittman17}. These systematic and geometric challenges render the translation of raw observables into a three-dimensional physical model a highly degenerate inverse problem. Addressing this degeneracy motivates the development of robust statistical frameworks, forward-modeling techniques, and~multi-wavelength synergy, which are actively being refined to enable the extraction of subtle particle physics signatures from current and next-generation~surveys.

Even when macroscopic offsets are observationally constrained, utilizing them to test fundamental dark matter physics requires rigorous comparison against numerical N-body hydrodynamical simulations, which introduce their own critical systematic limitations. The~most notable theoretical challenge concerns the gravitational softening length ($\epsilon_{\rm DM}$), a~strict numerical requirement in these models designed to prevent non-physical two-body scattering and shot noise at small scales \citep{Roche24}. Notably, the~application of this parameter intrinsically introduces an artificial flattening at the center of the simulated dark matter density profile. This ``artificial coring'' is mathematically degenerate with the physical coring that would be induced by actual SIDM, confounding the identification of genuine fundamental physics signatures within simulated halo~profiles.

Recent evaluations of state-of-the-art CDM cosmological simulations reveal a significant spatial convergence problem regarding these specific small-scale offsets. The~strict spatial scale required for numerical convergence in these models is typically defined as $2.8 \times \epsilon_{\rm DM}$ \citep{Roche24}. However, the~intrinsic median offsets measured between BCGs and their dark matter centroids in these simulations are frequently on the order of $\sim$1~kpc, which falls entirely below the resolution limit of even the highest-resolution cosmological runs~\citep{Roche24}. Extrapolating galactic dynamics and attempting to deduce fundamental dark matter behaviors at scales operating below the softening length yields fundamentally unconverged results. To~definitively test SIDM at the critical 1--10~kpc scale, future numerical campaigns will require significantly smaller softening lengths and rigorously proven convergence to properly complement the statistical ensembles derived from~observations.

\section{Conclusions}
\label{sec:conclusions}

Merging galaxy clusters rank among the most powerful macroscopic laboratories in the Universe for probing the fundamental particle physics of dark matter. Since the landmark observations of the Bullet Cluster, these cosmic colliders have unequivocally demonstrated the existence of the dark sector by revealing the macroscopic decoupling of the effectively collisionless mass distribution from the highly collisional baryonic gas. Beyond~merely confirming dark matter, these extreme dynamical environments have become indispensable for testing physics beyond the standard CDM paradigm. High-velocity mergers act as natural particle accelerators, providing direct empirical constraints on alternative frameworks, such as SIDM, and~bridging the critical gap between the largest and smallest scales known in~nature.

Multi-wavelength observational campaigns have significantly advanced our capacity to map the complex post-collision anatomy of these massive systems. By~synergizing X-ray maps of the ICM with gravitational lensing reconstructions of the dark matter potential, it is now possible to trace the dynamical evolution of these cosmic mergers with unprecedented detail. The~continually expanding catalog of observed systems marks a pivotal evolution in the field. Instead of relying solely on the deterministic analysis of unique, isolated cases, the~community is progressively leveraging the robust statistical power of heterogeneous merger ensembles, enabling a much broader understanding of dark matter decoupling across diverse collision~scenarios.

Realizing the full precision potential of these macroscopic colliders requires a continued, rigorous synthesis between expanding observational surveys and advanced theoretical models. Extracting definitive fundamental physics constraints relies on directly comparing measured spatial offsets against high-fidelity N-body hydrodynamical simulations. As~next-generation telescopes assemble statistically robust merger samples and computational frameworks achieve even greater precision, this multi-disciplinary synergy will reliably translate cosmological observables into fundamental constraints. By~fully integrating these vast empirical datasets with state-of-the-art numerical baselines, the~study of merging galaxy clusters holds the transformative potential to finally illuminate the true microscopic nature of the dark~sector.

\vspace{6pt}

\funding{This research received no external funding}

\dataavailability{Data sharing is not applicable to this article as no new data were created or analyzed in this study.}

\acknowledgments{I would like to express my deepest and warmest gratitude to the wonderful colleagues at the Academia Sinica Institute of Astronomy and Astrophysics (ASIAA), Taipei, Taiwan; the National Central University, Taoyuan, Taiwan; the National Institute of Physics, University of the Philippines Diliman; the Istituto Nazionale di Fisica Nucleare---Sezione di Torino, Italy; and the Center for Astrophysics and Cosmology at the Federal University of Espírito Santo, Brazil. The~concept for this review originated from a seminar they graciously hosted. Their welcoming spirit, generosity, and~stimulating discussions provided the true spark of inspiration for this work. Thank you all so deeply for your time and valuable insights!
During the preparation of this manuscript, the~author used Gemini Pro 3.1 for the purposes to improve the overall language of the manuscript. The~author has reviewed and edited the output and takes full responsibility for the content of this~publication.}

\conflictsofinterest{The author declares no conflicts of~interest.} 

\begin{adjustwidth}{-\extralength}{0cm}
\printendnotes[custom] 

\reftitle{References}

\PublishersNote{}
\end{adjustwidth}
\end{document}